\documentclass[format=sigconf,screen=true,nonacm=true]{acmart}

\usepackage[T1]{fontenc}
\usepackage{graphicx}
\usepackage{subcaption}
\usepackage{multirow}
\usepackage{bm}
\usepackage{tabularx}
\usepackage[table]{xcolor}
\usepackage{threeparttable}

\usepackage[shortcuts,acronym]{glossaries}
\glsdisablehyper
\newacronym[shortplural=GS]{gs}{GS}{Ground Station}
\newacronym{isl}{ISL}{Inter-Satellite Link}
\newacronym{ut}{UT}{User Terminals}
\newacronym{gsl}{GSL}{Ground-Satellite Link}
\newacronym{udp}{UDP}{User Datagram Protocol}
\newacronym{tcp}{TCP}{Transmission Control Protocol}
\newacronym{icmp}{ICMP}{Internet Control Messaging Protocol}
\newacronym{rtt}{RTT}{Round-Trip Time}
\newacronym[shortplural=DES]{des}{DES}{Discrete Event Simulation}
\newacronym{voip}{VoIP}{Voice over Internet Protocol}
\newacronym{lkm}{LKM}{Loadable Kernel Module}
\newacronym{bdp}{BDP}{Bandwidth-Delay Product}
\newacronym{cbr}{CBR}{Constant Bitrate}
\newacronym{vm}{VM}{Virtual Machine}
\newacronym{dce}{DCE}{Direct-Code Execution}
\newacronym{hil}{HIL}{Hardware-in-the-Loop}
\newacronym{tle}{TLE}{Two-Line Element}
\newacronym{qdisc}{qdisc}{queuing discipline}
\newacronym{leo}{LEO}{Low Earth Orbit}
\newacronym{csv}{CSV}{Comma-Separated Values}
\newacronym{ecn}{ECN}{Explicit Congestion Notification}
\newacronym{qos}{QoS}{Quality-of-Service}
\newacronym{mpi}{MPI}{Message Passing Interface}

\copyrightyear{2025}
\acmYear{2025}

\usepackage{listings}
\definecolor{background}{HTML}{F5F5F5}
\lstdefinelanguage{csv}{
    basicstyle=\footnotesize\ttfamily,
    stepnumber=1,
    showstringspaces=false,
    numbers=left,
    xleftmargin=1.5em,
    frame=single,
    framexleftmargin=1.5em,
    numbersep=0.5em,
    breaklines=true,
    frame=lines,
    backgroundcolor=\color{background},
    identifierstyle=\color{black},
    stringstyle=\color{black},
    captionpos=b,
    abovecaptionskip=0.6em,
    aboveskip=1.4em,
    belowskip=0.8em
}

\definecolor{DarkOrchid}{HTML}{9E508F}
\definecolor{WildStrawberry}{HTML}{FF3769}
\definecolor{Red}{HTML}{FF2E17}
\definecolor{Green}{HTML}{00AB4F}
\definecolor{RedOrange}{HTML}{FF5C31}
\definecolor{RoyalBlue}{HTML}{0073C0}

\begin{document}
\title{Trace-driven Path Emulation of Satellite Networks using Hypatia}

\author{Martin Ottens \href{https://orcid.org/0009-0003-4257-0087}{\includegraphics[scale=0.04]{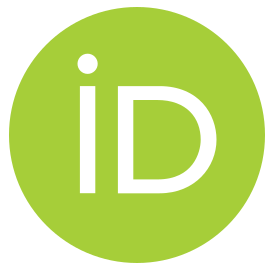}}{}}
\affiliation{%
  \institution{Friedrich-Alexander-Universität}
  \city{Erlangen-Nürnberg}
  \country{Germany}}
\email{martin.ottens@fau.de}

\author{Kai-Steffen Hielscher \href{https://orcid.org/0000-0002-2051-0660}{\includegraphics[scale=0.04]{figures/orcid.png}}{}}
\affiliation{%
  \institution{Friedrich-Alexander-Universität}
  \city{Erlangen-Nürnberg}
  \country{Germany}}
\email{kai-steffen.hielscher@fau.de}

\author{Reinhard German \href{https://orcid.org/0000-0002-9071-4802}{\includegraphics[scale=0.04]{figures/orcid.png}}{}}
\affiliation{%
  \institution{Friedrich-Alexander-Universität}
  \city{Erlangen-Nürnberg}
  \country{Germany}}
\email{reinhard.german@fau.de}

\renewcommand{\shortauthors}{Ottens et al.}

\begin{abstract}
  The increasing prevalence \ac{leo} satellite mega-constellations for global Internet coverage requires new approaches to evaluate the behavior of existing Internet protocols and applications.
  Traditional \acp{des} like Hypatia allow for modeling these environments but fall short in evaluating real applications.
  This paper builds upon our previous work, in which we proposed a system design for trace-driven emulation of such satellite networks, bridging the gab between simulations and real-time testbeds.
  By extending the Hypatia framework, we record network path characteristics, e.g., delay and bandwidth, between two endpoints in the network during non-real-time simulations. 
  Path characteristics are exported to Trace Files, which are replayed in real-time emulation environments on real systems, enabling evaluations with real software and human interaction.
  An advantage of our approach is its easy adaptability to existing simulation models.

  Our extensive evaluation involves multiple scenarios with different satellite constellations, illustrating the approach's accuracy in reproducing the behavior of satellite networks.
  Between full simulation, which serves as a baseline for our evaluation, and emulation runs, we observe high correlation metrics of up to $0.96$, validating the approach's effectiveness. 
  Challenges such as the lack of emulation-to-simulation feedback and synchronization issues are discussed.
\end{abstract}


\maketitle

\section{INTRODUCTION}
\label{sec:introduction}

The increasing amount of commercial and governmental projects involving \ac{leo} satellite mega-constellations to provide global Internet coverage~\cite{fortune2025} is currently sparking tremendous interest in this research area.
To evaluate various factors in the early project phase of a satellite network deployment, different types of simulation tools are employed~\cite{jiang2023}.
One aspect that needs to be assessed is the behavior of current Internet applications when transmission is handled via new mega-constellations.
Applications and protocols are often designed for terrestrial networks with relatively static characteristics.
Mega-constellation networks exhibit significantly more dynamic characteristics, e.g., due to frequent handovers between satellites or the complex routing within the constellation.
With the increasing prevalence of these networks, it is therefore necessary to evaluate the behavior of existing protocols, such as congestion control algorithms, and optimize them if required~\cite{lai2024}.
One approach to assess the performance and behavior of Internet protocols are \ac{des}.
Notable examples of \acp{des} extended for evaluating satellite mega-constellation networks are the ns-3-based \textit{Hypatia}~\cite{kassing2020} and \textit{ns-3-leo}~\cite{schubert2022} or the OMNeT++-based \textit{OS\textsuperscript{3}}~\cite{niehoefer2013}.

While \acp{des} are a common tool for evaluating simple Internet protocols, it becomes a challenge to evaluate complete applications with their protocol stacks, as simulation models of the applications are required, and/or no real-time interaction is possible with the applications under test during the simulation.
To circumvent these shortcomings, various approaches involving emulation testbeds, where a software emulates a link between real software components or even systems are proposed \cite{ohs2025,tian2024,kassem2024,ruan2025}.

\begin{figure}[t]
  \centering
  \includegraphics[width=\linewidth]{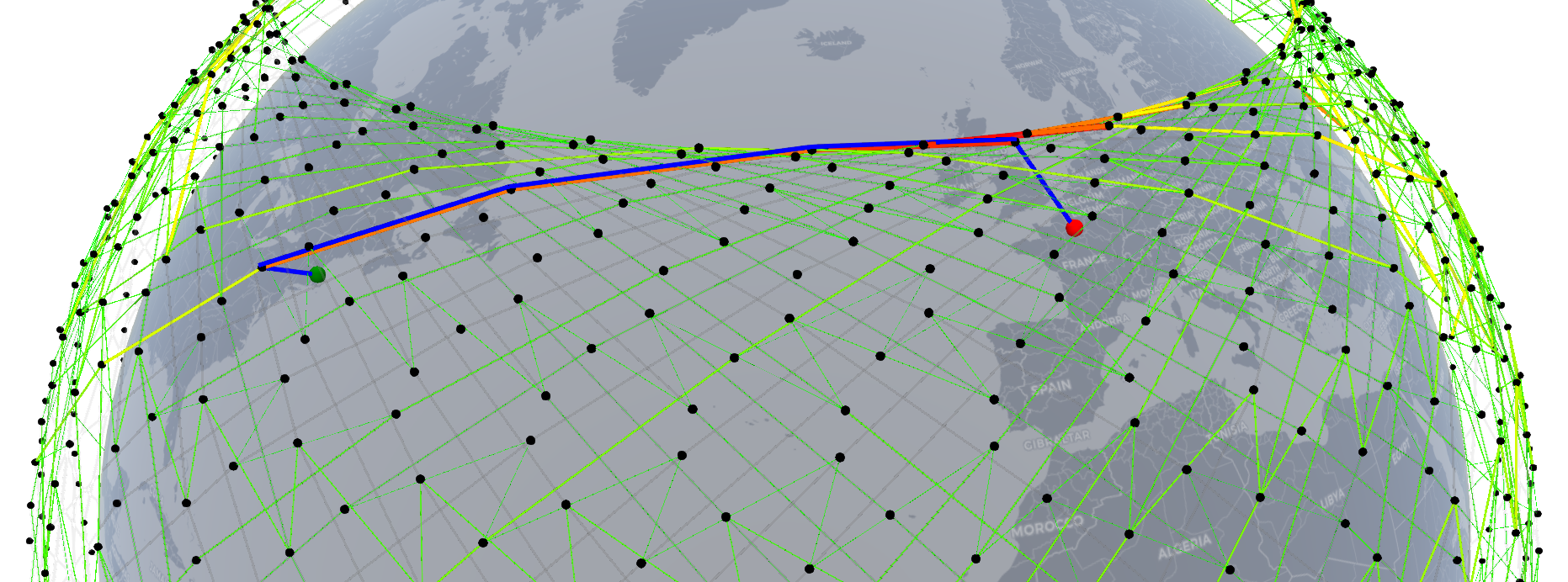}
  \caption{Visualization of the \acs{isl} load due to Background Traffic in a Starlink constellation (green = low \acs{isl} load, red = high \acs{isl} load) and the network path between a \acs{gs} in Boston and Paris (blue) at t=100s.}
  \label{fig:constellation_usage}
\end{figure}

To provide such an emulation testbed, we proposed a system design for a trace-driven emulation in our previous work \cite{ottens2025-2}.
This approach is detailed in Section~\ref{sec:trace_overview}.
A non-real-time simulation records a timeline of the varying network characteristics, such as delay, bandwidth, and queue capacity, of a path between two selected endpoints in the network and exports these characteristics to Trace Files.
These Trace Files are replayed in an emulation environment, where the path characteristics are applied to a link between two real systems communicating in real-time.
With our system design, we want to meet the following objectives:
\begin{itemize}
  \item It should be possible to implement complex routing algorithms and a traffic engineering system in the simulation. 
        The simulation should be able to handle large amounts of traffic, modeling global usage of the network.
  \item Current simulation frameworks are unable to provide this functionality in real-time; thus, the simulation should be able to run longer than real-time.
        Additionally, it should be possible to extend simulation models to generate Trace Files with minimal effort, allowing existing simulation setups to be utilized for this workflow.
  \item In the emulation environment, real software stacks should be used (e.g., real operating systems, application implementations, protocols).
        From the software stack's perspective, the path between the systems should provide the same characteristics as those that prevailed during the simulation.
  \item The emulation environment should run in real-time to allow human-in-the-loop interaction with the systems. It should be able to handle arbitrary traffic, e.g., different protocols, congestion control algorithms, larger traffic volumes, and bursty traffic.
\end{itemize}

In this paper, we ported our system design for trace-driven emulation to mega-constellation networks.
For this, we extended \textit{Hypatia}~\cite{kassing2020} with new functionalities and the ability to record network path characteristics to Trace Files.
With further improvements to our emulation environment, we demonstrate in a comprehensive evaluation that our system design is capable of reproducing the simulated behavior of such satellite networks.
As in our previous work, we used a full simulation as a baseline for our evaluation, where the network itself, as well as the test software is implemented as simulation models.
This baseline enables us to compare results from our emulation environment, where the test software runs on real systems and the Trace Files are replayed, with results from a deterministic simulation.

This paper is structured as follows: Section~\ref{sec:related_work} provides an overview of research on comparable emulation environments, emphasizing the differences from our approach.
Our simulation setup and the modifications to Hypatia are presented in Section~\ref{sec:simulation_setup}.
In Section~\ref{sec:trace_overview}, we provide an overview of the workflow and our emulation environment, highlighting the improvements to our previous work.
An evaluation that includes three different scenarios is presented in Section~\ref{sec:evaluation}.
Section~\ref{sec:real_time_mode} examines another approach for a real-time simulation that we investigated during our research, demonstrating its ineffectiveness for our objectives.
In Section~\ref{sec:discussion}, we discuss our evaluation results and the limitations of our approach.
Finally, Section~\ref{sec:conclusion} offers a conclusion and outlines directions for future work.

\section{RELATED WORK}
\label{sec:related_work}

In current research on the evaluation and improvement of Internet protocols via satellite mega-constellations, testbeds play an important role.
Often, real satellite networks or simple link emulations with static characteristics are used~\cite{hofstatter2025}.
Such testbeds are unable to provide reliable and reproducible results or make simplified assumptions about a network's behavior~\cite{lai2024}.
Some researchers also build more complex testbeds for their experiments~\cite{kosek2022}; a common tool for this is, for example, \textit{MiniNet}~\cite{mininet2022}.
Specific testbeds often remain unpublished or are complicated to set up, making reuse or later validation generally impossible.
For this reason, different works focus on developing general mega-constellation emulation environments with different approaches.

\textit{Xeoverse} by Kassem~et~al.~\cite{kassem2024} provides a two-stage real-time emulation environment.
In the first stage, satellite positions, delays, and routing tables are precomputed.
In a second stage, these results are applied continuously to a MiniNet network, which utilizes the network namespace and routing features of the Linux kernel to build a virtual network topology.
The first stage can calculate \ac{gsl} and \ac{isl} capacities based on weather data; however, it does not utilize the benefits of a \ac{des} to leverage simulation models for complex routing algorithms or to simulate other traffic in the network that would impact the selection and capacity of links.
The MiniNet environment is still elaborate and utilizes userspace code to apply link characteristic updates. 
Thus, the accuracy could be limited, and debugging such a setup it not trivial.
The two-stage approach is comparable to our system design, but our design leverages the benefits of a \ac{des} to generate the link characteristics for the emulation environment.
Other tools, such as \textit{StarryNet}~\cite{lai2023} and \textit{OpenSN}~\cite{lu2025}, use comparable virtual topologies that utilize Linux namespaces or containers.
Different to these approaches, our emulation environment focuses on a single end-to-end path as this often sufficient for protocol evaluation, eliminating the need for complex testbed topologies.

The emulation tool \textit{PhantomLink} by Ohs~et~al.~\cite{ohs2025} takes a similar two-stage approach and emulates a single end-to-end path.
Comparable to our approach, PhantomLink also uses \acs{csv} files to export the characteristics of the path from a simulation to the emulation environment.
However, it does not describe how to obtain these characteristics, whereas our system design encompasses the entire workflow.
Nevertheless, it could be possible to replay our Trace Files in PhantomLink.
PhantomLink uses a userspace program to perform the link emulation, probably a challenge for large data rates, and the system's configuration could limit the accuracy.

With \ac{dce}, ns-3 provides an option to run real software or even entire operating system kernels inside the simulation~\cite{camara2014}.
However, some software requires modification~\cite{dce2013}, and currently, support for \ac{dce} seems to have ceased.
Most importantly, by design, no real-time interaction with an application running in \ac{dce} is possible.

Another approach is to run a simulation framework in real-time and route traffic from real systems via the emulated topology.
ns-3 provides a real-time synchronization mode and TapBridges for this~\cite{nsnam2025-1}.
We have evaluated the performance of such setups in previous work~\cite{ottens2025-1}; a specific evaluation for satellite mega-constellation is summarized in Section~\ref{sec:real_time_mode}.
\textit{NSDocker} by Ruan~et~al.~\cite{ruan2025} also utilizes these features.
Therefore, comparable limits will apply, especially with complex topologies.
Haibin~Song~et~al. propose to parallelize the ns-3 simulation of large satellite constellations using \acs{mpi}~\cite{song2024}.
Due to the architecture of \acs{mpi}-based approaches, incorporating real-time emulation features with high accuracy is not possible, as these approaches are primarily designed for simulations.

In summary, our approach differs from others in that we present an entire workflow: from leveraging a \ac{des} simulation for the collection of path characteristics to evaluation in a real-time emulation environment.
Our approach for Trace Files generation can be easily adopted to other, existing simulator setups, whereby our emulation tooling is usable in a variety of different setups and can reach optimal accuracy.

\section{SIMULATION SETUP}
\label{sec:simulation_setup}

For this work, we updated parts of the open-source \ac{leo} simulation framework \textit{Hypatia}~\cite{kassing2020}\footnote{\url{https://github.com/snkas/hypatia/tree/master/ns3-sat-sim}} to ns-3 version \textit{3.44}.\footnote{\url{https://gitlab.com/nsnam/ns-3-dev/-/releases/ns-3.44}}
The newer ns-3 version incorporates several improvements and implements additional \acs{tcp} congestion algorithms, such as \textit{TCP CUBIC}~\cite{pan2022}, compared to the version \textit{3.31} that the published version of Hypatia uses.
After the update, we repeated all experiments from the original Hypatia paper to validate that the simulator still yields comparable results.
The updated simulator forms the basis for further extensions we implemented for our experiments.

\begin{table}[tb!]
\caption{Overview of the constellation configurations used in our experiments, as introduced in~\cite{kassing2020}.}
\label{tab:constellation_params}
\centering
\begin{tabular}{lccc}
\toprule
                                  & \textbf{Kuiper} & \textbf{Starlink} & \textbf{Telesat} \\ \midrule
\textbf{Height}                   & 630~km          & 550~km            & 1015~km          \\
\textbf{Orbits}                   & 34              & 72                & 27               \\
\textbf{Satellites per Orbit}     & 34              & 22                & 13               \\
\textbf{Inclination}              & 51.9\textdegree & 53\textdegree     & 98.98\textdegree \\
\textbf{Min. elevation}           & 30\textdegree   & 25\textdegree     & 10\textdegree    \\
\textbf{\acsp{isl} per Satellite} & 4               & 4                 & 4                \\
\textbf{Number of \acs{gs}}       & 100             & 100               & 100              \\ \bottomrule
\end{tabular}
\end{table}

\subsection{Satellite Constellations \& Network Topology}
\label{subsec:topology}
To minimize the number of newly introduced variables in our work to and to ensure comparability, we utilize the constellations and \ac{gs} configurations from the original Hypatia paper.
Due to Hypatia's design, it is possible to add simulations for additional configurations at a later stage.

Three different constellations are used in our experiments; an overview is provided in Table~\ref{tab:constellation_params}.
All three constellations have different orbits and heights for their satellites, which allows us to explore various effects in the networks.
The constellations use the same set of 100 \acp{gs}, which are located at populated places around the earth.
Communication happens between these \acp{gs}, so there are no dedicated \acp{ut} involved in the simulation scenarios.
Each satellite can maintain any number of \acp{gsl}, allowing simulation scenarios with \acp{ut} to be implemented at a later stage.
Each satellite has four \acp{isl}: Two to the next satellite on the same orbital plane and two to the nearest satellites on the neighboring orbital planes.
All \acp{isl} are established once the simulation is started, so there is no handover mechanism for the \acp{isl}.

Hypatia handles routing by calculating the network graph of the constellation at a specific interval (e.g., each 100~ms), where the distance is used as edge weights.
The all-pair shortest paths are then obtained using the Floyd-Warshall algorithm.
Each satellite and \ac{gs} has a routing table, defining which outgoing interface a packet for a specific destination must take.
These routing tables are called the \textit{Forwarding State}. 
Forwarding States are pre-calculated in a preparation step and exported to multiple files.
The ns-3 simulation reads these files and applies the updated Forwarding State to the routing tables of the satellites.
Therefore, routing decisions are made \textit{offline} before the simulation is executed -- the routing table update in the simulator is limited to the interval selected as an input of the preparation step.
Still, transmission times are calculated for each packet based on the exact satellite position at simulation-time.
The effects of this approach were investigated by the authors of Hypatia in their paper~\cite{kassing2020}.

To simulate the behavior of different network designs, we extended Hypatia with two mechanisms that can be enabled during the ns-3 simulations:
\begin{itemize}
  \item \textbf{Handover:} Hypatia implements \ac{gsl} handovers without packet loss.
        We added an option to drop all packets on a \ac{gsl} link for a configurable amount of time after a handover occurred.
  \item \textbf{Reconfiguration Interval:} In the Starlink network, a global synchronization is done in a 15-second interval during which no packets are transmitted via \acs{gsl}~\cite{mohan2024}.
        In Starlink, these global synchronizations are used to reconfigure the network: For example, handovers are handled during this reconfiguration.
        We added an option that stops the transmission of packets for a configurable amount of time in a configurable interval.
        In contrast to packet drops during handovers, packets (e.g., produced by applications running on a \ac{gs}, but also the Trace Packets used to obtain the path characteristics) are not dropped; instead they are enqueued for later transmission if sufficient queue capacity is available.
\end{itemize}
Another extension is a simple traffic engineering mechanism, which reserves a certain bandwidth for selected \acp{gsl} for specific traffic.
\acp{isl} are not affected by this traffic engineering mechanism, allowing to observe the effects of congestions within the network.

In addition to these extensions, other parameters such as \ac{isl} and \ac{gsl} link bandwidths and queue capacities can be selected for each ns-3 simulation run.
All settings together allow for an easy comparison of different constellation designs and their scalability in the future.
To demonstrate the capabilities of our approach, we present different scenarios with three satellite constellations and different configurations in this paper.

\subsection{Background Traffic}
\label{subsec:background_traffic}

The scope of our work is to analyze the effects of the dynamic link characteristics of a path between two specific endpoints (e.g., two \acp{gs}) on real protocols and applications.
Since the bandwidth of \acp{gsl} and \acp{isl} in a commercial satellite constellation is likely shared, the crosstraffic, i.e., the traffic generated by other users within the network, can have a significant impact on the path characteristics and possibly also on routing decisions.
In the following, this crosstraffic is referred to as \textit{Background Traffic} due to its peculiarities.

Other works, such as those by Roth~et~al., have incorporated complex methods to model the traffic exchanged in satellite constellations in their simulations~\cite{roth2022}.
However, in our case, we selected a more simplified approach: 10,000 flows are exchanged during the simulation between the 100 \acp{gs}, where the duration, bandwidth, as well as the source and destination \acp{gs}, are pseudo-randomly selected.
The definition for all flows is exported to a file that is replayed during ns-3 simulations using a \acs{udp} sender model provided Hypatia.
All simulation scenarios in our experiments encounter the same Background Traffic for comparability.
The parameters are manually selected so that all constellations show different conditions during a single simulation run.
Parameters for the Background Traffic are summarized in Table~\ref{tab:background_params}.

\begin{figure}[tb!]
  \centering
  \includegraphics[width=\linewidth]{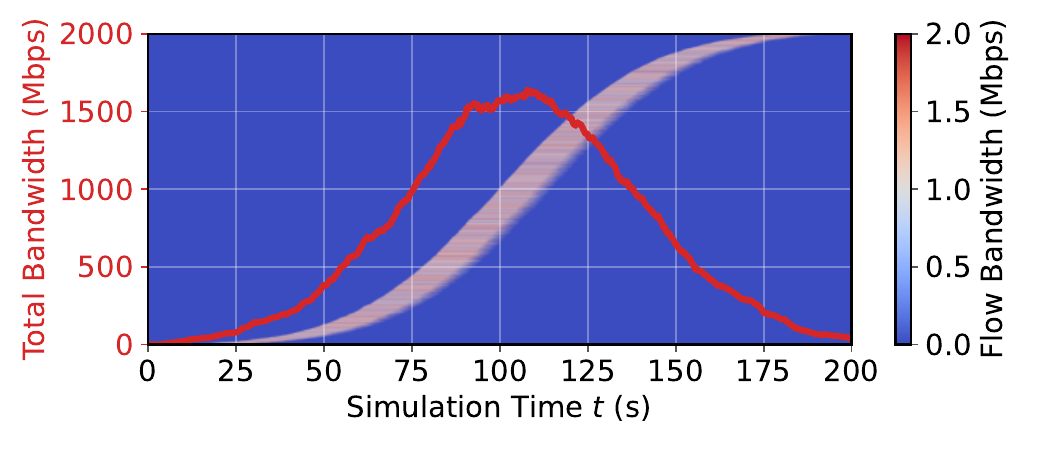}
  \caption{Visualization of the Background Traffic flows and their total bandwidth during the 200-second simulation.}
  \label{fig:background_traffic}
\end{figure}

Since we are only running \textit{transient} simulations~\cite{wehrle2010} that terminate after 200 seconds, the active number of flows is normally distributed over the simulation time.
In our case, the largest amount of background traffic in the network is present at around $t=100 s$.
This enables us to observe changes in the network's behavior under different load conditions.
The flows for the Background Traffic used for all our scenarios, as well as the total bandwidth, are visualized in Figure~\ref{fig:background_traffic}.
The resulting load caused by the background traffic on the \acp{isl} of the Starlink constellation at $t=100 s$ can be seen in Figure~\ref{fig:constellation_usage}.

As described in our previous work, \acs{udp} traffic is used for the Background Traffic, as it is not affected by congestion control algorithms~\cite{ottens2025-2}.
During a full simulation, other traffic types are not problematic; however, during emulation, the lack of feedback from the emulation environment back to the Trace File generation in the simulation could falsify the measurements, as detailed in Section~\ref{sec:discussion}.

\begin{table}[tb!]
\caption{Parameters for Background Traffic generation.}
\label{tab:background_params}
\centering
\begin{tabular}{lc}
\toprule
\textbf{Parameter}       & \textbf{Value}     \\ \midrule
Total Number of Flows    & 10000              \\
Bandwidth per Flow       & 0.1 to 2.0~Mbps    \\
Duration per Flow        & 10 to 15~s         \\
\bottomrule
\end{tabular}
\end{table}

\section{TRACE-DRIVEN EMULATION OF SATELLITE NETWORKS}
\label{sec:trace_overview}

Simulations of complex satellite networks are time-consuming, often making it impossible to execute the simulation in real-time.
The authors of Hypatia and ns-3-leo investigated the runtime of their simulation: With large simulated traffic volumes or complex routing algorithms, the runtime exceeds the simulated time by many times~\cite{kassing2020,schubert2022}.
For this reason, we have proposed a novel approach for trace-driven emulation in previous work~\cite{ottens2025-2}. 
In this approach the link characteristics of the path through a network topology packets would take between two endpoints are recorded in a simulation.
These characteristics are exported to \textit{Trace Files}, which can be replayed in an emulation environment in real time, where the recorded path characteristics are applied.
This approach decouples the time-consuming simulation from the real-time emulation environment, allowing for a seamless integration in testbeds.

\subsection{Workflow Overview}
\label{subsec:workflow_overview}

\begin{figure}[tb!]
  \centering
  \includegraphics[width=\linewidth]{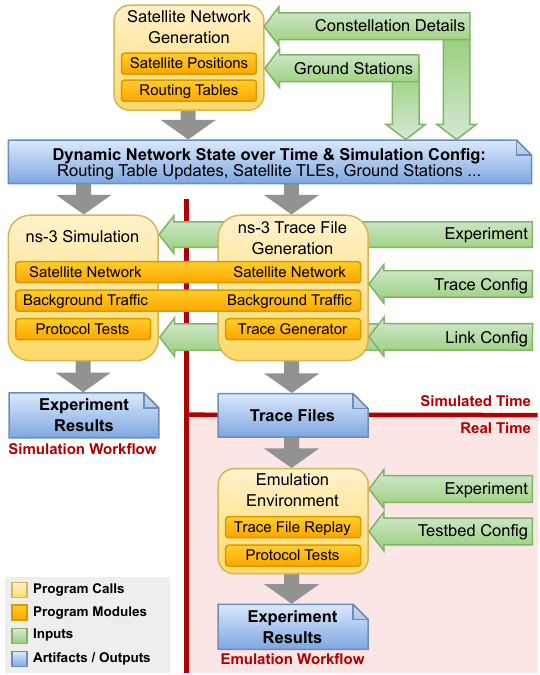}
  \caption{Overview of the workflows used in this paper.
           The left side shows the workflow of a full simulation, the unmodified Hypatia workflow. The right side shows the emulation workflow with the additional real-time step added.}
  \label{fig:trace_sim_workflow}
\end{figure}

Our previous work details the workflow of this approach for a simple topology.
In this work, we ported our previous Trace-File generation implementation to Hypatia.
Figure~\ref{fig:trace_sim_workflow} provides an overview of the resulting workflow.

The first step in a Hypatia simulation is to calculate the \textit{network dynamic state over time}, which is a timeline of changes in the routing tables for all satellites and \acp{gs}.
During this step, Hypatia receives a list of \acp{gs} and the satellite constellation definition as input, generating routing table updates at configurable intervals (e.g., every 100~ms).
Also, \acp{tle} for the satellites are generated during this step.

After this initial step, the \ac{des} is performed.
The simulator takes the data generated in the first step and additional configuration parameters (e.g., the maximum bandwidth of \acp{isl} and \acp{gsl}) as an input to perform various experiments.
During the simulation, all transmission delays between satellites and/or \acp{gs} are calculated based on the exact satellite positions at the time of transmission; only the routing table updates are limited to the update interval selected in the first step.
In the case of a typical full simulation, which is the workflow the authors of Hypatia have used in their work, this step yields the results from the experiments for further analysis.

In our extended workflow, the actual experiments are not performed in this step; instead, it is only used to generate the Trace Files.
Besides generating the Background Traffic and application of the pre-generated routing table updates, which both are also present during a full simulation, the simulator takes care of measuring the network characteristics of the path between two pre-selected \acp{gs}.
During this step, the update resolution for the Trace File is selected, i.e., the interval in which updates of the path characteristics are recorded.
Multiple Trace Files can be generated simultaneously, but they always need to be generated as a pair: one file for the \textit{forward} direction (\ac{gs} $A \rightarrow B$) and another for the \textit{return} direction (\ac{gs} $B \rightarrow A$).
These Trace Files can be archived, since, as long as no input parameters of the simulation change, they can be used for multiple experiments in the emulation environment.

In our workflow, the actual experiments take place in a third step.
In this step, the Trace Files are replayed in real-time between two systems, emulating the path between them.
The systems host the applications or protocols that are tested in real-time, and user interaction with these systems is possible.

In comparison to our previous work, a new parameter is exported to the Trace Files: As proposed by Ohs~et~al.~\cite{ohs2025}, a \textit{Route Identifier} is calculated to allow distinguishing between different routes of the end-to-end path.
Packets transmitted in the emulation environment while the same route identifier is active cannot implicitly reorder (e.g., due to a high delay jitter or when the delay fluctuates heavily between characteristic updates).
We assume that packets will not recorder within the same route of a real satellite network; only when the route changes, e.g., due to a satellite handover, reordering is possible.

\subsection{Kernel Module \& Emulation Environment}
\label{subsec:emulation_environment}
The emulation environment is based on a Linux \ac{vm} that can be connected to the other systems by physical links or virtual network bridges~\cite{liu2022}.
In our previous work, \textit{NetEm}~\cite{hemminger2005} was used on the system together with scripts running in userspace, which allowed for a satisfactory reproduction of path characteristics with a Trace File update resolution of 100~ms.
Since applications running in userspace can be delayed significantly by the kernel's scheduler, link characteristic updates may be applied with severe delay, which affects the accuracy and reproducibility of the setup.
Especially in satellite networks, a 100~ms update resolution might not be sufficient to reproduce the short-term effects of highly dynamic networks.
In Section~\ref{subsec:sync_interval}, we demonstrate a use case that could be impacted by a too coarse update resolution.

To enable more precise and deterministic path characteristic update, we developed \textit{TheaterQ}, an \ac{lkm} that provides a \ac{qdisc} that can be installed at every egress interface.
TheaterQ has comparable functionality to NetEm, but can update settings (bandwidth, delay, queue capacity, packet loss) without support from the userspace once it is set up.
Since TheaterQ provides more features that can make it suitable for a variety of different use cases, we published the source code along with a detailed documentation\footnote{\url{https://github.com/cs7org/TheaterQ}}.

\begin{figure}[tb!]
  \centering
  \includegraphics[width=\linewidth]{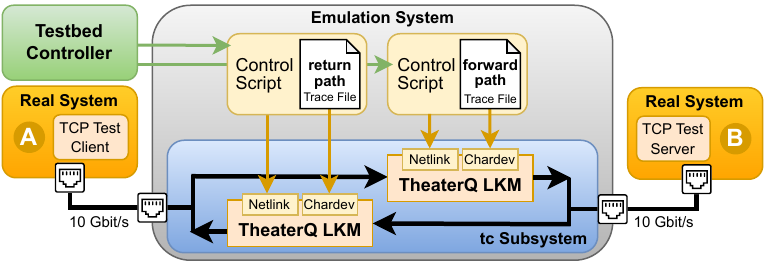}
  \caption{The emulation setup is conceptually similar to the one presented in~\cite{ottens2025-2}.
           It consists of three \acp{vm} managed by a testbed controller.
           Traffic is routed through an emulation system, where a \ac{lkm} replays the link characteristics from the Trace Files.}
  \label{fig:emulation_setup}
\end{figure}

Figure~\ref{fig:emulation_setup} shows how TheaterQ is integrated into our emulation setup.
Two TheaterQ instances are used: one works on the egress of the interface to system $A$, handling with the forward path's Trace File, while the other, on egress of the interface to $B$, handles the return path.
For controlling the TheaterQ instances, additional scripts are running.
The scripts translate the Trace Files before ingestion via the character devices and are no longer involved in the emulation as soon as the replay in the \ac{lkm} is started.

Our testbed consists of three \acp{vm}: Two for the endpoints, running the software under test, and one as a dedicated emulation system, running TheaterQ.
All \acp{vm} are using \textit{Debian 12} with kernel \textit{6.1.0-37}.
The entire testbed, including the control scripts, is managed via an external testbed controller~\cite{ottens2025-3}.
The testbed controller also handles the temporal synchronization of all script calls and manages the software under test installed on the endpoint systems.
It is worth mentioning that TheaterQ and our workflow can also be used in other setups, such as a \textit{MiniNet} environment or with simple namespace or container-based setups.

\section{EVALUATION}
\label{sec:evaluation}

To evaluate the accuracy of our trace-driven emulation approach for satellite networks, we conducted several experiments with different scenarios.
Our software under tests consists of a simple \acs{icmp} ping and a \acs{tcp} speedtest tool, comparable to \textit{iperf3}~\cite{iperf3}.
However, our speedtest tool can measure additional metrics like transmission times and congestion windows for individual packets.
The following two different workflows are compared:
\begin{itemize}
  \item \textbf{Full Simulation:} The \acs{tcp} speedtest and \acs{icmp} ping, our software under test, is implemented as a simulation model.
        This workflow is shown in Figure~\ref{fig:trace_sim_workflow} on the left side: After generating the satellite network state, all other tests take place in the ns-3 simulation.
        Since the simulation is deterministic, the results are reproducible. 
        This is the native Hypatia workflow and serves as the baseline for the evaluation.
  \item \textbf{Emulation:} This workflow is shown in Figure~\ref{fig:trace_sim_workflow} on the right side: The ns-3 simulation does not run the software under test; only the Trace Files are generated in the simulator.
        The Trace Files are used in the emulation environment after the simulation is completed, where the actual software under test is executed in real-time.
\end{itemize}
In both workflows, the satellite network and the Background Traffic is simulated. 
The main difference lies in when the software under test is executed.

In the following three scenarios are presented, an overview is shown in Table~\ref{tab:evaulation_overview}. 
All scenarios use some or all of the satellite constellations described in Section~\ref{subsec:topology} and the Background Traffic introduced in Section~\ref{subsec:background_traffic}.
The scenarios were selected to enable a performance comparison between the satellite constellations under different configurations.
Although this is not the primary objective of this work, we intend to utilize our workflow for this very purpose in the future.
Parameters for Background Traffic volume and link data rates were selected so that our environment can demonstrate its ability to reproduce different network behaviors.

Our \ac{vm}-based emulation environment introduces additional delays to the \acs{rtt} due to drivers and bridge interfaces.
These delays are static for the most part, allowing them to be hidden by offsetting the delay values from the Trace Files.

\begin{table}[t!]
\caption{Overview of the evaluation scenarios.}
\label{tab:evaulation_overview}
\centering
\begin{tabularx}{\columnwidth}{lccc}
\toprule
\textbf{}                                                             & \textbf{Scenario 1}                                                 & \textbf{Scenario 2}                                         & \textbf{Scenario 3}                                                       \\ \midrule
\textbf{Constellations}                                               & \begin{tabular}[c]{@{}c@{}}Kuiper\\ Starlink\\ Telesat\end{tabular} & Kuiper                                                      & Kuiper                                                                    \\ \hline
\textbf{Ground Stations}                                              & \begin{tabular}[c]{@{}c@{}}Boston\\ Paris\end{tabular}              & \begin{tabular}[c]{@{}c@{}}Boston\\ Paris\end{tabular}      & \begin{tabular}[c]{@{}c@{}}Rio de Janeiro\\ St. Petersburg\end{tabular} \\ \hline
\textbf{Trace File shown}                                                   & Starlink                                                            & Kuiper                                                      & Kuiper                                                                    \\ \hline
\textbf{Measurements}                                                 & \begin{tabular}[c]{@{}c@{}}Goodput\\ Ping RTT\end{tabular}          & \begin{tabular}[c]{@{}c@{}}Goodput\\ Ping RTT\end{tabular}  & Ping RTT                                                                  \\ \hline
\textbf{\begin{tabular}[c]{@{}l@{}}Additional\\ Aspect\end{tabular}} & \begin{tabular}[c]{@{}c@{}}Handover\\  Packet Loss\end{tabular}     & \begin{tabular}[c]{@{}c@{}}Network\\ Reconfigs\end{tabular} & \begin{tabular}[c]{@{}c@{}}Connection\\ Loss\end{tabular}                \\ \bottomrule

\end{tabularx}
\end{table}

\subsection{Scenario 1: Different Satellite Constellations}
\label{subsec:compare_constellations}

This scenario compares the achievable \acs{tcp} goodput and \acs{icmp} ping \acs{rtt} between \acp{gs} in Boston and Paris using three different satellite constellations.
The parameters used for the ns-3 simulation are summarized in Table~\ref{tab:scenario_params}.
Each \ac{gsl} and \ac{isl} has a maximum data rate of 50~Mbps and a queue capacity of 200~packets. 
Whenever a \ac{gsl} handover occurs, packets are dropped for 250~ms.
The Forwarding State and, thus, the routing tables of the satellites and \ac{gs} are updated every 100~ms.

When the ns-3 simulation is used to generate Trace Files, the virtual Trace Packets are transmitted every 2~ms. 
Trace Packets are \textit{virtual}, as they do not have a transmission time and therefore have no effect on link utilization.
Trace Packets can be enqueued in each interface queue along the transmission path.
However, they are excluded from the computation of the queue capacity, as only real traffic occupies link capacity and queue space in the simulation.
To account for short fluctuations and to calculate a coarse packet drop ratio, the average of five subsequent Trace Packets is used, resulting in a Trace File update resolution of 10~ms.
Such a coarse drop ratio is sufficient, as our experiment configuration drops packets for multiple milliseconds; therefore the drop ratio is 1 or 0 in most cases.

\begin{table}[tb!]
\caption{Parameters for full simulation and Trace File generation with scenario-specific configurations.}
\label{tab:scenario_params}
\centering
\begin{tabularx}{\columnwidth}{lc}
\toprule
\textbf{Parameter}            & \textbf{Value}                        \\ \midrule
Routing Table Update Interval & 100~ms                                \\
Data Rate per ISL / GSL       & 50~Mbps / 50~Mbps                     \\
Queue Size per ISL / GSL      & 200~pkts / 200~pkts                   \\
Trace Packet Interval         & 2~ms                                  \\
Average of $n$ Trace Packets  & 5 $=$ 10~ms update resolution         \\ \midrule
\textit{Scenario 1:}          &                                       \\
Packet loss during Handover   & 250~ms                                \\ \midrule
\textit{Scenario 2:}          &                                       \\
Reconfiguration Interval      & 15~s                                  \\
Reconfiguration Duration      & 100~ms                                \\
\bottomrule
\end{tabularx}
\end{table}

\subsubsection{Trace File}
\begin{figure*}[tb!]
  \centering
  \includegraphics[width=\linewidth]{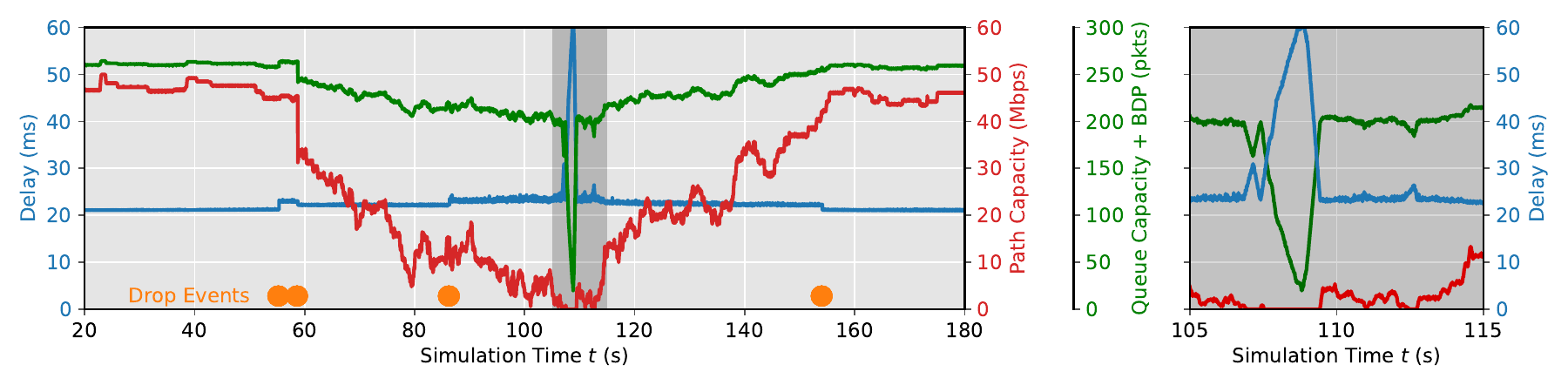}
  \caption{\textit{Scenario 1:} Selected properties from the Trace File of the path between Boston and Paris using the Starlink constellation.
          Orange dots indicate temporary packet losses that occurred during \ac{gsl} handovers in this scenario.
          The area on the right shows the time between 105 and 115~seconds in detail, where a congested \ac{isl} caused packets to accumulate in the queue before the link.}
  \label{fig:scenario_1_trace}
\end{figure*}

A visualization of selected values of an exported Trace File is shown in Figure~\ref{fig:scenario_1_trace}, an excerpt from the Trace File in a \ac{csv} format can be found in Listing~\ref{lst:trace_file}.
The figure shows the characteristics of the forward path between Boston and Paris when the Starlink constellation is used. 
Another Trace File contains the path characteristics for the return path, e.g., the path from Paris to Boston.
This scenario, with the \ac{isl} load due to Background Traffic in the constellation at $t=100 s$, is also shown in Figure~\ref{fig:constellation_usage}.
The small delay changes and drop events in Figure~\ref{fig:scenario_1_trace} indicate a \ac{gsl} handover.
In this scenario, packet loss only occurs during handovers in all three satellite constellations.
In Figure~\ref{fig:scenario_1_trace}, it is evident that the available path and queue capacity decreases with increasing Background Traffic load (cf. Figure~\ref{fig:background_traffic}).
Since the network is highly dynamic, with flows starting and stopping and satellite movements triggering handovers, there are significant fluctuations in both available path bandwidth and queue capacity.

We selected this Trace File from a Starlink constellation, as a link in the path is fully congested at around $t=108 s$ (cf. the dark area and magnification on the right side of Figure~\ref{fig:scenario_1_trace}).
This is because the selected path runs via the upper envelope of the constellations in the polar region and there is no traffic engineering or routing in the simulation that tries to move traffic to alternate paths. 
During this time, packets accumulate in the queue of the interface before the congested link, lowering the available queue capacity and increasing the delay.
No packet drop occurs, since the queue can be emptied fast enough after a path change in this case.
This is the behavior we expected and shows that our Trace File generation is capable of recording such complex interactions.

\begin{table*}[htb!]
\begin{threeparttable}
\caption{\textit{Scenario 1:} Statistical comparison of results from the full simulation and emulation using the average of 25 repeated runs in the emulation environment.
         The reproducibility between multiple emulation results is also shown for the \acs{tcp} speedtests.
         Lag-corrected values are calculated by offsetting the timing differences between different runs with an optimal fit.}
\label{tab:scenario1_stats}
\centering
\begin{tabularx}{\textwidth}{lccc|ccc|ccc}
\toprule
                                     & \multicolumn{3}{c|}{\textbf{Emulations vs. Emulations}}  & \multicolumn{6}{c}{\textbf{Full Simulation vs. Emulations}}                                                     \\ \midrule
                                     & \multicolumn{3}{c|}{\textbf{\acs{tcp} Goodput}}               & \multicolumn{3}{c|}{\textbf{\acs{tcp} Goodput}}               & \multicolumn{3}{c}{\textbf{\acs{icmp} No-load \acs{rtt}}}          \\
                                     & \textbf{Kuiper} & \textbf{Starlink} & \textbf{Telesat} & \textbf{Kuiper} & \textbf{Starlink} & \textbf{Telesat} & \textbf{Kuiper} & \textbf{Starlink} & \textbf{Telesat} \\ \midrule
\textbf{Mean Average Error}          & 1.18 Mb/s       & 1.42 Mb/s         & 4.8 Mb/s         & 3.76 Mb/s       & 3.77 Mb/s         & 4.3 Mb/s         & 0.58 ms         & 0.52 ms           & 0.57 ms          \\
\textbf{R²}                          & 0.93            & 0.91              & 0.67             & 0.79            & 0.86              & 0.56             & 0.88            & 0.81              & 0.85             \\
\textbf{Correlation (Pearson)}       & 0.97            & 0.95              & 0.84             & 0.92            & 0.94              & 0.68             & 0.86            & 0.82              & 0.89             \\ \midrule
\textbf{Average Lag}                 & 0.02 s          & 0.09 s            & 0.19 s           & 0.95 s          & 1.05 s            & 1.00 s           & 1.02 s          & 1.07 s            & -\tnote{a}                \\
\textbf{Correlation (Lag-corrected)} & 0.98            & 0.96              & 0.89             & 0.95            & 0.96              & 0.68             & 0.91            & 0.94              & -\tnote{a}                \\ \midrule
\textbf{Total Transmission}          & 505 MB          & 514 MB            & 680 MB           & 505 MB          & 514 MB            & 680 MB           & -               & -                 & -                \\
\bottomrule
\end{tabularx}
{
  \footnotesize
  \begin{tablenotes}
  \item[a]{No features in results that allowed for an accurate lag calculation.}
  \end{tablenotes}
}
\end{threeparttable}
\end{table*}

\subsubsection{TCP Goodput}
\label{sec:scenario_1_tcp_goodput}
\begin{figure*}[tb!]
  \centering
  \includegraphics[width=\linewidth]{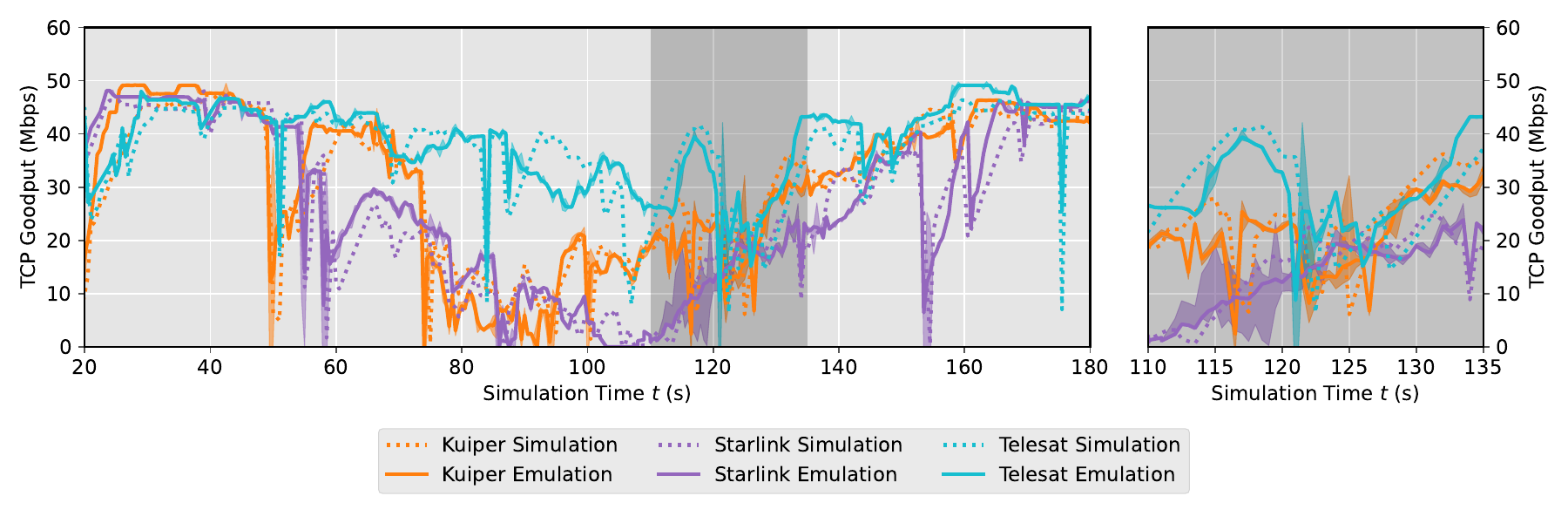}
  \caption{\textit{Scenario 1:} \acs{tcp} goodput measured in simulations and emulations of three different satellite constellation scenarios.
          The first and last 20~seconds are omitted to visualize the goodput during steady-state, e.g., without inaccuracies during setup.
          The area on the right shows the time between 110 and 135~seconds in more detail.
          The shaded areas in the emulation results represent the range between the 10th and 90th quantiles, based on 25 different repetitions.
          The static lines are an average value from these repetitions.}
  \label{fig:scenario_1_result_goodput}
\end{figure*}

Figure~\ref{fig:scenario_1_result_goodput} compares the \acs{tcp} goodput of the three different constellations measured in a full simulation and the emulation environment replaying the Trace Files.
For the goodput measurements, custom and comparable \acs{tcp} speedtest implementations were used in the emulation and as a simulation model.
These implementations were also used in our previous work.\footnote{See \url{https://github.com/cs7org/Trace-File-based-Network-Emulation-Demo} for both implementations.}
\textit{TCP CUBIC} was used as the congestion control algorithm.
On the emulation systems, default \acs{tcp} settings were used, while the \acs{tcp} parameters in the full simulation were configured to closely match those of the emulation environment.

Nevertheless, the \acs{tcp} and network stack implementations in a real Linux kernel and ns-3 differ significantly~\cite{prakash2020}; therefore, we expect the results to differ to a certain extent in both cases.
It is also worth mentioning that the emulation hosts run in real-time; thus, the kernel can delay some transmissions.
In contrast, the full simulation with a fixed seed always yields the same, reproducible results~\cite{nsnam2025-2} for a given scenario, which is a key advantage of a \ac{des}.

In difference to our previous work, the simulated satellite constellations are much more dynamic than simple network topologies.
Therefore, the achievable throughput encounters high fluctuations in all cases.
Figure~\ref{fig:scenario_1_result_goodput} shows that the \acs{tcp} goodput achieved in the emulations is close to our baseline, the full simulations, most of the time.
Differences are evident when the path capacities are near zero or when the available path capacity changes rapidly.
Rapid changes result in short over- and undershoots in the \acs{tcp} goodput, as shown in the magnified area of Figure~\ref{fig:scenario_1_result_goodput}.

\begin{figure*}[tb!]
  \centering
  \includegraphics[width=\linewidth]{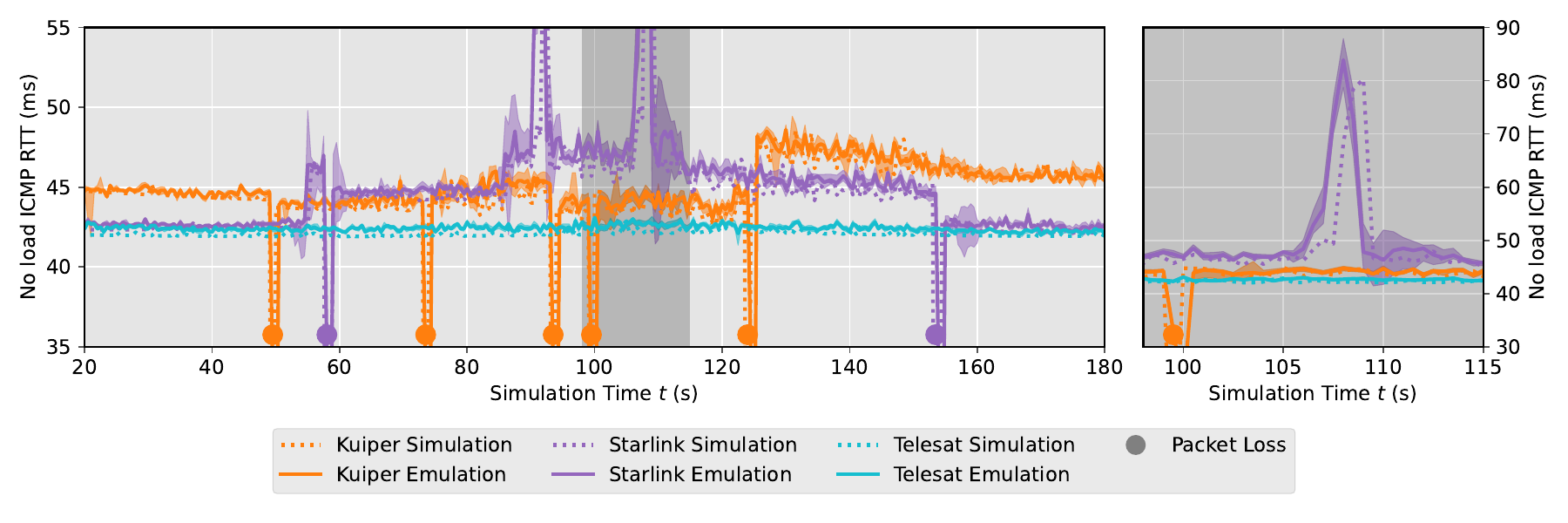}
  \caption{\textit{Scenario 1:} \acs{icmp} ping \acs{rtt} measured without parallel \acs{tcp} flows, but with Background Traffic in simulations and emulations of three different satellite constellations.
          The area on the right shows the time between 98 and 115~seconds in more detail with a different y-axis scaling, showing the Starlink congestion recovery from Figure~\ref{fig:scenario_1_trace}.
          The shaded areas in the emulation results represent the range between the 10th and 90th quantiles, based on 25 different repetitions.
          The static lines are an average from these repetitions.}
  \label{fig:scenario_1_result_ping}
\end{figure*}

Since there is no randomness in our full simulation, repeated runs, even with different seeds, will yield the same measurements and Trace Files.
Our emulation environment is based on real \acp{vm}, where multiple involved schedulers introduce randomness; thus, repeated runs will not yield identical results.
To assess the ability of our approach to predictably reproduce the behavior of a network path and its accuracy across multiple experiments, we conducted 25 independent tests in our emulation environment.
Table~\ref{tab:scenario1_stats} presents a statistical comparison of these results, and Figure~\ref{fig:scenario_1_result_goodput} shows the 10\% and 90\% quantiles of these runs alongside the average as a shaded area.
As expected, a difference between multiple emulation runs is present, but it is minimal.
It is worth mentioning that the results from the Telesat emulation have the most fluctuations.
In this case, changes to the delay and bandwidth are relatively small compared to the other emulations, allowing the non-deterministic effects of the schedulers to become more visible.
There are some differences between the results from the emulation and full simulation, e.g., the emulation environments overestimate the bandwidth by around 7\%.
These differences correlate with the total amount of traffic that was transmitted during the \acs{tcp} speedtests.

An important aspect is the synchronization between the starting time of the test software, in our case, the \acs{tcp} speedtest or \acs{icmp} pings, and the replay of the Trace Files.
To investigate this, we measured the lag between multiple emulation runs and between the emulations and the simulation.
Between emulations, the differences are minimal, but our emulation environment lags around 1 second behind the simulation. 
When correcting this offset, the correlation between emulation and full simulation increased accordingly.
A static offset can correct these effects, but this offset highly depends on the specific emulation environment used for the experiments and needs to be measured during the setup.

Summarized, the differences in between full simulation and emulation in Figure~\ref{fig:scenario_1_result_goodput} and~\ref{fig:scenario_1_result_ping} are minimal.
Differences between multiple different measurements in real networks are often much greater.
We are not aware of any other approach that is capable of reproducing such a detailed network behavior in a real-time emulation environment with better accuracy.
We selected synthetic \acs{tcp} test traffic as this allows for a comparison with a full simulation.
Tests with real applications and traffic are subject to future work.
Also, the performance analysis of different satellite constellations can be done in future work. 
Interesting effects are, for example, that the Starlink constellation, which uses the highest number of satellites, shows a reduced throughput due to \ac{isl} congestion, especially when compared to the smaller Telesat constellation. 

\begin{figure*}[h!]
  \centering
  \includegraphics[width=\linewidth]{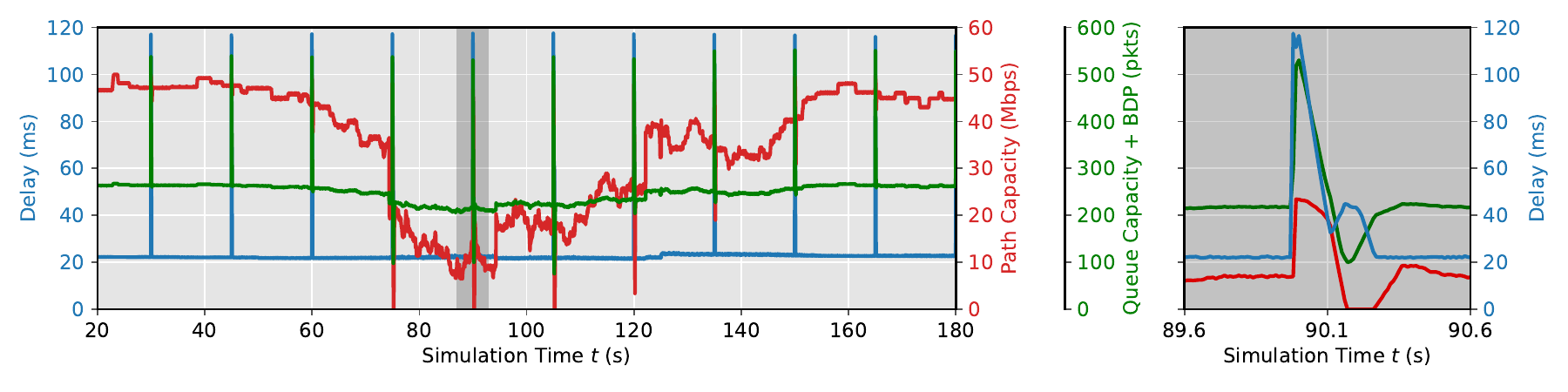}
  \caption{\textit{Scenario 2:} Selected properties from the Trace File of the path between Boston and Paris using the Kuiper constellation with a 15-second global reconfiguration interval.
          During a reconfiguration, all \acp{gsl} stop transmitting packets.
          The area on the right of the figure shows such a reconfiguration between 89.6 and 90.6~seconds in detail.}
  \label{fig:scenario_2_trace}
\end{figure*}

\subsubsection{RTT}
Another important aspect is the reproduction of the delay along the network path.
Different applications, such as \ac{voip}, typically have low data rates but are susceptible to changes in the delay.
To demonstrate the ability of our emulation environment to reproduce the delays from the simulation, we compared the delays of the three constellations in Figure~\ref{fig:scenario_1_result_ping}.
\acs{icmp} pings were used to measure the \ac{rtt} with \textit{iputils} ping, which is also implemented in ns-3.
To avoid interference caused by the different behavior of the network stack and qdics in ns-3 and Linux, the ping was measured without a parallel \acs{tcp} flow.
We refer to this as \textit{No-load \acs{rtt}}, as the \ac{gsl} of the endpoint \acs{gs} was not loaded to capacity; nevertheless, the Background Traffic was present during the simulations, and the \acsp{gs} exchanging the pings were also involved in sending and receiving the Background Traffic.
Pings were sent every 0.5~seconds.

Figure~\ref{fig:scenario_1_result_ping} shows that the emulation environment closely reproduces the delays encountered during the full simulation.
The magnified area of this Figure shows another challenge: in case of the Starlink constellation, it is visible that the emulation encountered a higher \ac{rtt} of 0.5 seconds before this was expected in the full simulation.
This is because it is impossible to precisely determine when a ping is sent in such emulation environments.
Also, the exact start point of the \ac{lkm} Trace File replay, which is controlled from within the userspace, is not perfectly deterministic.
Both are limitations of a real-time emulation environment, and comparable experiments via real networks would encounter the same limitations.

Table~\ref{tab:scenario1_stats} also shows a comparison between the \acs{rtt} in the full simulation and multiple emulation runs.
During the emulation of all three constellations, the results are comparable: the emulation overestimates the \acs{rtt} on average by around 500~µs, which is a difference of approximately 1\%.
In our emulation environment, additional delays from the setup were compensated with a static delay offset, which must be selected identically for each emulation setup prior to experiments.
The lag already highlighted in Section~\ref{sec:scenario_1_tcp_goodput} is also present in this case and can be corrected with another static offset, further improving the accuracy of the emulation setup.
Nevertheless, the emulation environment is capable of reproducing delays with a satisfactory accuracy, especially between different emulation runs an average correlation $\ge 0.98$ allows for good reproducibility.

\subsection{Scenario 2: Global Network Reconfiguration}
\label{subsec:sync_interval}

The second scenario is comparable to the Kuiper setup in Scenario 1: the path between \ac{gs} in Boston and Paris is used and the same Background Traffic is present.
The difference is that \acs{gsl} handovers are handled without packet loss.
Instead, the network uses the reconfiguration intervals, comparable to those present in the real Starlink network.~\cite{mohan2024}
Reconfigurations occur every 15~seconds.
During such a reconfiguration, all \ac{gsl} interfaces stop transmitting packets for 100~ms, but no packets are dropped as long as sufficient queue capacity is available.

\begin{figure*}[h!]
  \centering
  \begin{subfigure}{.33\textwidth}
    \centering
    \includegraphics[width=.99\linewidth]{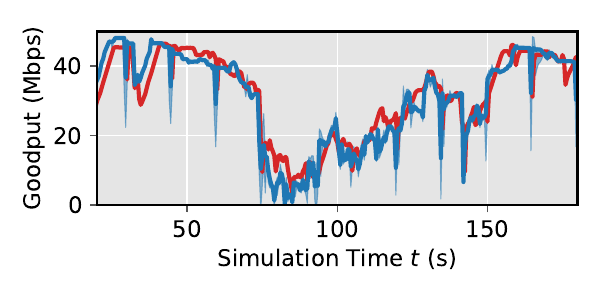}
    \caption{\acs{tcp} goodput during speedtest}
    \label{fig:scenario_2_result_goodput}
  \end{subfigure}%
  \begin{subfigure}{.33\textwidth}
    \centering
    \includegraphics[width=.99\linewidth]{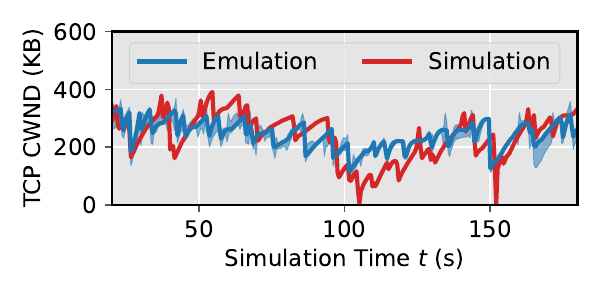}
    \caption{Congestion Window during speedtest}
    \label{fig:scenario_2_result_cwnd}
  \end{subfigure}%
  \begin{subfigure}{.33\textwidth}
    \centering
    \includegraphics[width=.99\linewidth]{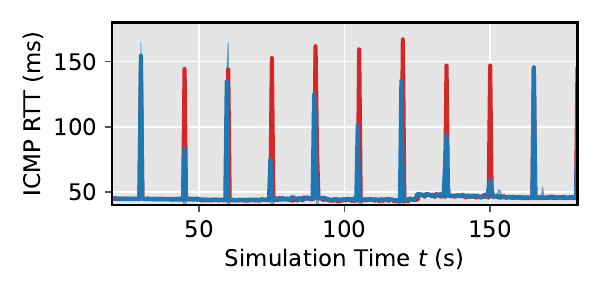}
    \caption{No-load \acs{icmp} \acs{rtt}}
    \label{fig:scenario_2_result_ping}
  \end{subfigure}%
  \caption{\textit{Scenario 2:} Comparison of different metrics recorded in the full simulation and the emulation of the Kuiper constellation between \acp{gs} in Boston and Paris with 15-second global reconfiguration intervals.
           The shaded areas in the emulation results represent the range between the 10th and 90th quantiles, based on 25 different repetitions.
           The blue line is an average from multiple emulations runs.}
  \label{fig:scenario_2_result}
\end{figure*}

\begin{figure*}[h!]
  \centering
  \includegraphics[width=\linewidth]{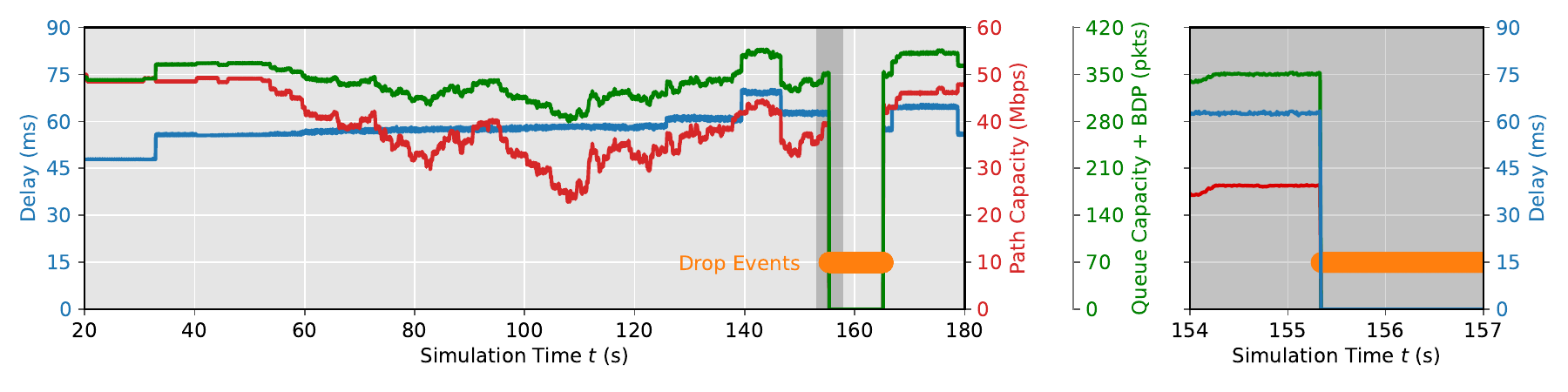}
  \caption{\textit{Scenario 3:} Selected properties from the Trace File of the path between Rio de Janeiro and Saint Petersburg using the Kuiper constellation.
          Orange dots indicate temporary packet losses that occurred due to a too low elevation of the satellite in Saint Petersburg.
          The area on the right shows the time between 154 and 157~seconds in detail where the temporary connection dropout begins.}
  \label{fig:scenario_3_trace}
\end{figure*}

\subsubsection{Trace File}
As shown in Figure~\ref{fig:scenario_2_trace}, the Trace File records the same behavior during all reconfigurations.
The magnified area in the figure shows a particularly noteworthy effect when the global reconfiguration occurs while the constellation is highly loaded with Background Traffic.
The Trace File continuously records the capacity of the link with the least amount of available bandwidth a virtual Trace Packet encounters while it is transmitted via the current path. 
Also, the queue of the interface just before this bottleneck link is used to track the queue capacity.
The sending timestamp of the Trace Packets is used for the corresponding values recorded in the Trace Files.

When the reconfiguration starts at $t=90 s$, all Trace Packets, as well as packets from the Background Traffic, are not transmitted for at least 100~ms and are stored in the queues of the \ac{gsl} interfaces.
We selected the parameters of this scenario to ensure that the investigated network path can handle the reconfigurations without any packet losses.
After the reconfiguration is completed, Trace Packets are transmitted through the network alongside the Background Traffic that has been accumulated in the queues.
The effects of this can be seen at $t=90 s$: The available path capacity jumps as the bottleneck changes after the transmission starts again.
Since the Trace Packets were buffered in the \ac{gsl} interface, they record a higher path capacity at this time, since the \ac{isl} interfaces still transmitted during the reconfiguration, moving the position of the network's bottleneck.
The delay increases, as expected, to a maximum of 120~ms (100~ms reconfiguration plus the previous delay of the constellation of around 20~ms).
In case a Trace Packet is buffered in a queue, the end-to-end delay increases.
The queue capacity in the Trace File visualization also contains the \ac{bdp} of the full path to account for transmission and queuing delays.
As the available path capacity and, in particular, the delay increase, the \ac{bdp} rises sharply.
The queue capacity alone just accounts for the queue before the bottleneck interface.
Not accounting for other queues in this case would lead to unexpected packet losses.
However, since the network was capable of buffering more packets in the simulation during this scenario, as the increase in end-to-end delay without packet losses shows.

During normal operation of the constellation, packets do not accumulate in the queue, since no significant congestion occurs. During the 100~ms reconfiguration, packets start to accumulate in the queue before the bottleneck.
At $t=90.1 s$, the global reconfiguration is completed, and the constellation proceeds with normal operation.
Now, all accumulated packets are transmitted as fast as possible.
Another link becomes the bottleneck of the path, and since this link is fully utilized (with zero capacity), packets start to accumulate again at the affected interface queue, thereby reducing the path capacity and increasing the end-to-end delay.
At around $t=90.35 s$, the network has fully recovered from the reconfiguration: the queue backlog is fully processed.

\subsubsection{Measurements}
The \acs{tcp} goodput in Figure~\ref{fig:scenario_2_result_goodput} essentially shows the same effects and accuracy as described in Scenario 1 in Section~\ref{subsec:compare_constellations}.
This is also the case for the statistical insights described in this section.
During reconfiguration, it is likely that \textit{TCP CUBIC} will overload the queue of the \ac{gsl} interface; therefore, congestion events occur, and the congestion control algorithm reduces the congestion window.
This behavior can be clearly observed in both the full simulation and the emulation, although some differences are also present due to variations in timing and implementation differences.

Significantly larger differences are observed in the actual congestion window during the speedtest, as shown in Figure~\ref{fig:scenario_2_result_cwnd}.
We also observed this in our previous work, which used a much simpler network topology, and we assume that this is due to differences in the network stack implementations of ns-3 and the Linux kernel.
Since the goodput is still very comparable and reductions during congestion events are visible, these discrepancies should be negligible.

Figure~\ref{fig:scenario_2_result_ping} shows the no-load \acs{icmp} ping \acs{rtt}.
In the emulation environment, the exact timing of when ping requests are sent out is not controllable.
Since all effects of the reconfiguration only last a maximum of 0.35~seconds and pings are transmitted every 0.5~seconds, the exact timing when the ping is sent is relevant for the observed \acs{rtt}.
In this case, the additional one-way delay for a packet transmission can be $\in [0, 0.35] s$, depending on the exact timing.
Nevertheless, all reconfigurations are visible within the figure, and there is no unwanted overshooting of the \ac{rtt}.
This means for \ac{cbr} applications that a roughly comparable amount of packets is delayed during a reconfiguration, but it is not deterministic which individual packets of a flow are affected.
Once again, this drawback is also present in experiments with real networks and can be mitigated by running multiple repetitions.

\subsection{Scenario 3: Connection Loss}
\label{subsec:connection_loss}

The third scenario was chosen to demonstrate the capabilities of our approach in reproducing the effects of a longer connection dropout.
For this scenario, the Kuiper constellation is used; handovers are handled without packet loss, and there are no global reconfigurations.
The path between a \ac{gs} in Rio de Janeiro and Saint Petersburg is observed.
The Background Traffic is identical to that in the previous two scenarios.

\subsubsection{Trace File}
Due to the high inclination of the Walter Delta orbits of the Kuiper constellation, the high minimum elevation of the constellation, and the northern location of Saint Petersburg, the \ac{gs} loses connection to all satellites at around $t=155.4 s$.
This dropout is visible in Figure~\ref{fig:scenario_3_trace} and lasts for approximately 10~seconds.
During this time, the \ac{gsl} interface drops all packets; therefore, the Trace Packets are marked as lost, and no metrics for the delay, path, or queue capacity are recorded.
As soon as a new satellite is in reach again, the Trace Files contain these values again, as transmission are successful again.

\subsubsection{RTT}
\begin{figure}[tb!]
  \centering
  \includegraphics[width=\linewidth]{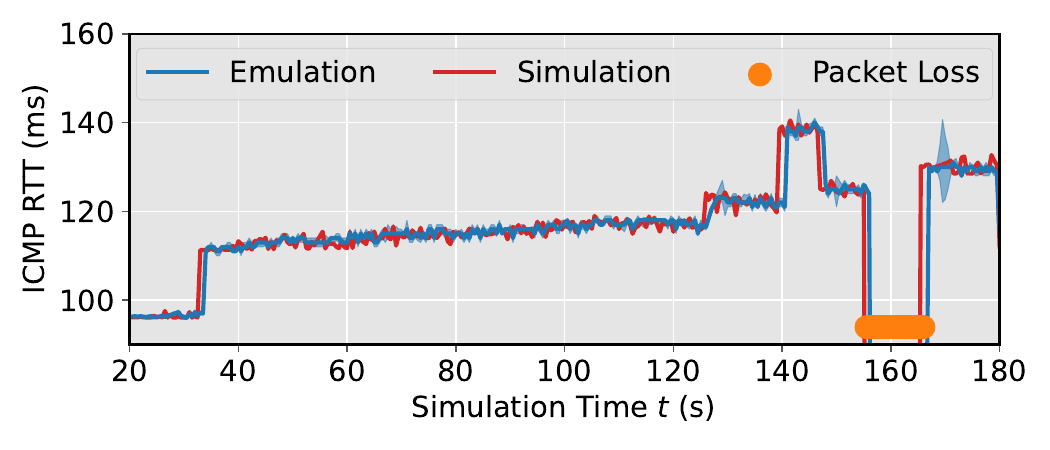}
  \caption{\textit{Scenario 3:} Comparison of the no-load \acs{icmp} \acs{rtt} in a full simulation and multiple emulations.
           The shaded areas in the emulation results represent the range between the 10tn and 90th quantiles, based on 25 different repetitions, the blue line is the average value.}
  \label{fig:scenario_3_result_ping}
\end{figure}

Figure~\ref{fig:scenario_3_result_ping} compares the \acs{rtt} of a no-load \acs{icmp} ping in a full simulation and emulation.
Since the ping implementations are identical in the Linux emulation environment and ns-3, and the processing of \acs{icmp} packets without additional load is comparable in both environments, these experiments yield very similar results.
As expected, all pings during the dropout fail until a new satellite comes into reach.

Additionally, the delay changes due to satellite handovers are clearly visible.
For an update of the delay in the emulation environment every 10~ms, the average delay of five Trace Packets during the Trace File generation in the simulation is used.
For this reason, the emulation curve in Figure~\ref{fig:scenario_3_result_ping} appears slightly smoother than the simulation, which records the exact \ac{rtt} during packet transmission.

Besides the lag of around 1~ms, the accuracy of the emulation environment is very high in this case.
The statistical values for the reproducibility between a full simulation and emulation, as well as between multiple emulations, are comparable to the values described in Section~\ref{sec:scenario_1_tcp_goodput}.
Comparable results could be obtained for the \acs{tcp} goodput measurements.

\section{NS-3 REAL-TIME EMULATION}
\label{sec:real_time_mode}

\begin{table}[tb!]
\caption{Selected results from tests with ns-3's real-time mode using 25 repetitions.
         \textit{E/S} compares the real-time mode emulation results with the simulation results, \textit{E/E} compares the reproducibility of repeated real-time mode emulation runs.
         $\bm{MAE}$ is the mean average error, $\bm{r}$ is the Pearson correlation (lag-corrected).}
\label{tab:real_time_emulation_results}
\centering
\begin{tabularx}{\columnwidth}{lll}
\toprule
\textbf{Testcase}                                                                        & \textbf{Sync Mode}                                         & \textbf{Result}                                                                                       \\ \midrule
\multirow{2}{*}{\begin{tabular}[c]{@{}l@{}}TCP Speedtest,\\ BG Traffic\end{tabular}}     & \begin{tabular}[c]{@{}l@{}}HardLimit \\ 1.0 s\end{tabular} & Simulation aborted                                                                                    \\ \cline{2-3} 
                                                                                         & BestEffort                                                 & Simulation aborted                                                                                    \\ \midrule
\multirow{2}{*}{\begin{tabular}[c]{@{}l@{}}TCP Speedtest, \\ No BG Traffic\end{tabular}} & \begin{tabular}[c]{@{}l@{}}HardLimit \\ 1.0s\end{tabular}  & Simulation aborted                                                                                    \\ \cline{2-3} 
                                                                                         & BestEffort                                                 & \begin{tabular}[c]{@{}l@{}}E/S: $MAE=19.5 \text{Mb/s}$, $r=0.32$\\ E/E: $MAE=0.94 \text{Mb/s}$, $r=0.60$\end{tabular} \\ \midrule
\begin{tabular}[c]{@{}l@{}}ICMP Ping,\\ BG  Traffic\end{tabular}                         & BestEffort                                                 & Simulation aborted                                                                                    \\ \midrule
\begin{tabular}[c]{@{}l@{}}ICMP Ping, \\ No BG Traffic\end{tabular}                      & \begin{tabular}[c]{@{}l@{}}HardLimit\\ 0.1 s\end{tabular}  & \begin{tabular}[c]{@{}l@{}}E/S: $MAE=1.9 \text{ms}$, $r=0.56$\\ E/E: $MAE=1.3 \text{ms}$, $r=0.69$\end{tabular}      \\ \bottomrule
\end{tabularx}
\end{table}

In previous work, we have investigated ns-3's real-time emulation capacities~\cite{ottens2025-1} using its TapBridge feature~\cite{nsnam2017}.
Even when useful for simple network topologies with a limited volume of real traffic, ns-3's real-time mode cannot cope with large and complex networks, and its accuracy decreases with higher traffic volumes or particularly bursty traffic.

\begin{figure}[tb!]
  \centering
  \includegraphics[width=\linewidth]{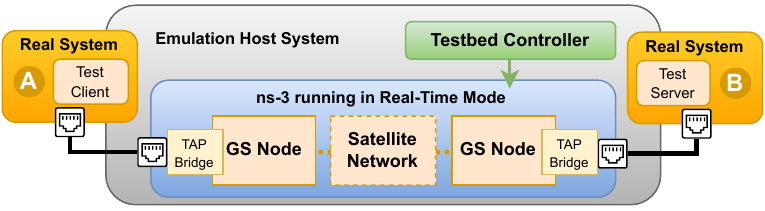}
  \caption{Emulation setup using ns-3's real-time mode with TapBridges.}
  \label{fig:real_time_emulation_setup}
\end{figure}

Regardless of our previous insights, we have modified Hypatia to enable its use in such a real-time mode emulation setup.
Our test setup is shown schematically in Figure~\ref{fig:real_time_emulation_setup}.
Comparable to the Trace File generation setup, a pair of \acp{gs} is selected prior to the start of ns-3.
ns-3 adds TapBridge NetDevices to the simulated \ac{gs} nodes and adapts the routing tables accordingly; the rest of the simulation remains unchanged.
Bridge interfaces are used on the Linux host to route the real traffic from the \acp{vm} via the TAP interfaces created by the simulation process.
ns-3 provides two modes for its real-time scheduler~\cite{nsnam2025-1}:
\begin{itemize}
  \item In the \textbf{HardLimit} synchronization mode, the simulation clock is only allowed to lag a limited amount of time (e.g., 100~ms) behind real time.
        When the lag becomes too large, the simulation process is terminated, which could happen during a large routing update or when a burst of real application traffic has to be processed.
  \item In the \textbf{BestEffort} synchronization mode, no such limit is given.
        This allows the simulator to run slower than real-time to some extent, especially as it can catch up again after a short traffic burst.
        There is no guarantee for the accuracy of the simulation in this mode, real packets can be delayed in a non-deterministic way.
\end{itemize}
We have tested both modes in our evaluation, but we only consider ns-3's HardLimit synchronization mode, with a strict limit, as applicable for future emulation setups.
All tests were conducted with the Kuiper constellation from \textit{Scenario 1} (cf. Section~\ref{subsec:compare_constellations}) with the optimized ns-3 build profile on recent hardware\footnote{CPU: Intel Xeon Silver 4310 (Mid 2021; 12~Cores, 24~Threads @ $2.10$-$3.30$GHz)}.

An overview of selected results from the real-time mode emulation is shown in Table~\ref{tab:real_time_emulation_results}.
In real-time mode, the simulator is unable to generate the Background Traffic without exceeding its execution time constraints.
This can also be seen from the fact that a non-real-time simulation of the scenario for 200~seconds with Background Traffic takes around 13.5~minutes of real time.

The simulation process is unable to handle the high packet throughput during the \acs{tcp} speedtest with a HardLimit of 1~second.
With the BestEffort synchronization mode, the simulation can handle the \acs{tcp} speedtest traffic, but the results of the speedtest do not reflect the behavior of the \acs{tcp} speedtest in a full simulation.
Additionally, the differences between multiple real-time mode emulation runs are substantial, indicating that the simulation process is overloaded and limiting the applicability of this simulation setup for larger traffic volumes.

\begin{figure}[tb!]
  \centering
  \includegraphics[width=\linewidth]{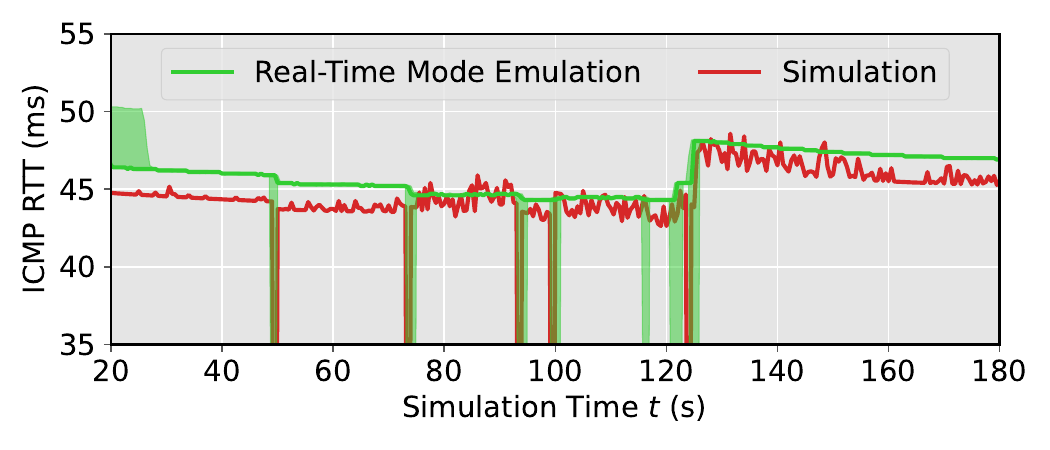}
  \caption{No-load \acs{icmp} \acs{rtt} in the Kuiper constellation of \textit{Scenario 1} in a simulation compared with a ns-3 real-time mode emulation.
           The shaded areas of the emulation results represent the range between the 10th and 90th quantiles, based on 25 repetitions.
           The green line is an average from all repetitions.}
  \label{fig:real_time_emulation_result}
\end{figure}

In the case of \acs{icmp} pings without Background Traffic, the real-time mode emulation achieves results with limited accuracy, as shown in Figure~\ref{fig:real_time_emulation_result}.
In this figure, the same effects of synchronization inaccuracies between the test systems and the trace-driven emulation as described in Section~\ref{subsec:connection_loss} are visible.
We applied the static delay offset from the previous emulation tests; however, the real-time mode emulation still overestimates the \acs{rtt} by around 1~ms.
All static offsets could be corrected during the setup, but this approach in general yields an inferior accuracy compared to our trace-driven emulation approach.

Since our \acs{tcp} speedtest, which can only reach up to 50~Mbit/s and around 6000~pps, is enough to overwhelm the simulation process with a 1-second HardLimit, scalability is a problem in such a real-time mode emulation.
The simulation process cannot be scaled with modern multi-core systems, as its core is working single-threaded.
In the case of ns-3, current work describes approaches to allow for multi-thread simulations~\cite{bai2024}, but it is unclear whether this approach can work accurately for real-time mode emulations.

\section{DISCUSSION \& LIMITATIONS}
\label{sec:discussion}

Our evaluation shows that our workflow and emulation environment are, by and large, capable of reproducing the behavior of single \acs{tcp} flows and \acs{icmp} ping \acsp{rtt} to that obtained in a simulation environment.
There are some differences in the behavior, but these need to be put into perspective.
Most differences can be explained by variations in timing and implementation of the deterministic models used in the full simulation and the real Linux network stack used in the \acp{vm} of the emulation environment.
Notably, the differences observed during multiple runs of the emulation environment are attributed to minor timing variations.
In comparison to evaluations performed with real networks, as common practice in research, our approach is still capable of reliably reproducing events that would only appear by chance in a real network.
Such events could be short connection dropouts due to obstructions, or congestions in the network.
We believe that using such an emulation environment could be a better choice for protocol research.
Since our emulation environment is based on a kernel module, accuracy will not be reduced as much when larger traffic volumes need to be processed, e.g., for testing a high-definition video meeting, as it could be the case with userspace emulation tools from other approaches.

As discussed in our previous work~\cite{ottens2025-2}, our approach has some limitations: the traffic encountered during emulation cannot affect the simulation, thus changing the routing decisions or altering the behavior of crosstraffic during congestions.
This limitation affects, for example, the interference between multiple \acs{tcp} congestion control algorithm instances.
After an infinite amount of time, two \acs{tcp} flows that attempt to achieve maximum throughput via a shared bottleneck would, by design, fairly share the bandwidth of that bottleneck.
During a simulation, the bottleneck would be fully used; in this case, the Trace File would contain a near-zero path capacity in this case.
A third identical \acs{tcp} flow in the simulation would get a third of the bottleneck bandwidth and reduce the goodput of the other two flows accordingly.
When this third flow appears in the emulation, the Trace File replay sets the path bandwidth to this near-zero value.
As a result, the third \acs{tcp} flow cannot achieve any throughput.

During the simulations in this paper, we used \acs{udp} flows with fixed bandwidths for the crosstraffic, as it is not affected by congestion control.
Basic traffic engineering measures that reserve a portion of the \ac{gsl} bandwidth of the endpoints for the \acs{tcp} speedtest or the Trace File generation ensured that congestion only appears on \acp{isl}.
Especially during light congestion, the interactions between \acs{udp} and \acs{tcp} flows will still be measurable to some extent when packet loss appears; we selected the congestion in our scenario to prevent that.
Real, commercial satellite networks would implement more complex traffic engineering measures that should prevent that traffic from different users from triggering such interferences.
Such mechanisms need to be implemented in the simulation itself.
With large constellations and high traffic volumes, the bandwidth required for a connection on the path between the selected endpoints may be negligible, allowing traffic engineering to reserve a specific bandwidth for that connection.
The missing feedback from emulation back to simulation is also a challenge for routing algorithms and networks that rely on \acp{ecn}~\cite{rfc3168}.
Future work could investigate whether mechanisms could reproduce \acp{ecn} in the emulation environment.
For the evaluation of the fairness of different \ac{tcp} congestion control algorithms it is of course possible to run multiple flows in parallel in the emulation environment.
Part of the Background Traffic is no longer handled in the simulation, but moved to the real-time emulation, where interferences between different flows are possible.

In our previous work, we described that full-duplex channels, e.g., a link where the return and forward paths share the available bandwidth, cannot be handled by our approach.
Such channels are uncommon in current satellite networks, as different frequencies are used for up- and downlinks~\cite{laniewski2024}.
However, future work could extend the Trace File generation as well as the kernel module, allowing for the emulation of such channels, which could be useful for other areas of network research.

In this work, we modified Hypatia to demonstrate that our system design can be employed to emulate dynamic satellite networks.
Due to Hypatia's two-step approach, the routing decisions were made \textit{offline}, before the simulation is started and cannot be altered during the simulation, e.g., due to local congestion.
To fully leverage the abilities of the \ac{des}, routing algorithms should be executed during the simulation, as done with ns-3-leo~\cite{schubert2022}, for example.
Better simulations could also be achieved by improving the Background Traffic.
For instance, more realistic traffic distributions, such as those used by Roth~et~al.~\cite{roth2022}, could be employed to better investigate the effects of paths between different geographic locations.

Our evaluation shows that our emulation environment yields very reproducible results between different runs with the same Trace Files.
However, during our experiments, we encountered a challenge that may affect this reproducibility: The synchronization between the emulation tool and the software under test.
Especially during the development of new protocols or modifications, it may be necessary for effects in the Trace Files to appear during a specific state of a protocol, e.g., during the slow start phase of a \acs{tcp} connection.
For this, the protocol test must start with a precise time difference to the start of the path emulation.
TheaterQ has an additional mode, in which the first packet handled triggers the replay of the Trace File, which can help align the emulation with the software under test in regard to timing.
To get a better understanding of the behavior of our emulation environment, we have not used this mode during our evaluation.
We plan to further enhance our emulation environment with synchronization features, allowing researchers to simply supply their Trace Files and software, and everything else is handled by the emulation environment in the background.
However, since we just aim to provide an emulation environment, we need to differentiate from full simulations. 
A time-perfect alignment and perfect reproducibility are not the objectives, as we rely on real network stacks that will always introduce some kind of randomness.

\section{CONCLUSION \& FUTURE WORK}
\label{sec:conclusion}

In this work, we extended our previous work on a system design for trace-driven emulations to allow an emulation environment to reproduce the network characteristics obtained from a satellite mega-constellation simulation.
Trace Files, containing path characteristics, are obtained from a modified version of Hypatia.
To accurately replay these Trace Files in an emulation environment, we presented TheaterQ, a kernel module providing a \ac{qdisc} that replays different network characteristics with accuracy and short update intervals, without involving the userspace.
We also provided insights into the workflow that researchers can utilize to create such Trace Files and use them in emulation environments for evaluating real protocol implementations or to conduct real-time application tests.
An important benefit is, that existing simulation setups can be modified to export Trace Files -- there is no need to re-develop a suitable simulation from scratch, as demonstrated in this work with Hypatia.

To evaluate our approach, we created three different scenarios involving different satellite constellations and configurations.
We used the results from a \acs{tcp} speedtest and \acs{icmp} pings as metrics for our evaluation.
A full simulation, where the speedtest and ping are implemented as simulation models, serves as a baseline.
The evaluation shows that our approach is capable of reproducing the effects of packet drops during satellite handovers, longer connection losses, as well as short local congestions in the network.
Between the results from a simulation and multiple emulation runs, we achieved correlations of up to $0.96$ with the \acs{tcp} goodput from our speedtest and up to $0.94$ with \acs{icmp} pings.
Between the results from different emulation runs, we achieved correlations up to $0.98$ with the \acp{tcp} goodput measurements.
During the analysis of our emulation results, it becomes evident that the synchronization between the different components involved in the workflow presents a challenge: minor differences in start times of components can result in lag, which reduces the reproducibility of experiments.

Finally, we discussed some limitations of our approach and why many of them are negligible when emulating satellite mega-constellations.
We provided basic ideas on how some current limitations can be solved in future work.
We will also continue to improve the usability of our workflow in the future by addressing the following points:
\begin{itemize}
  \item Modelling a network that is capable of simulating the behavior of the Starlink network and exporting corresponding Trace Files. 
        It might be possible to obtain models able to approximate the characteristics of the Starlink networks without detailed knowledge of its internal workings, e.g., using detailed analysis of the network's behavior~\cite{garcia2023}.
        This would allow us to directly compare the results of our emulation environment with measurements in the real Starlink network.
  \item To allow the evaluation of networks that use different paths for \ac{qos} traffic classes, we would like to extend our workflow to record multiple Trace Files for these paths during the simulation.
        An extended emulation environment could be capable of routing the traffic through different links with different characteristics.
  \item In the long term, we want to provide an easy-to-deploy usability testbed: An emulation computer between real systems should allow stakeholders of a satellite project to directly feel and compare how applications react when used via satellite constellations with different configurations.
        For this, a mature standard for Trace Files and a comprehensive software suite is required.
        A file standard could also allow interoperability with other tools, e.g., PhantomLink.
\end{itemize}

In the future, we plan to further improve this approach and provide a valuable tool for the development and modifications of Internet protocols that can keep pace with the demands of future satellite Internet connectivity.

%

\balance

\bibliographystyle{ACM-Reference-Format}
\bibliography{bibliography}


\begin{thebibliography}{36}


\ifx \showCODEN    \undefined \def \showCODEN     #1{\unskip}     \fi
\ifx \showISBNx    \undefined \def \showISBNx     #1{\unskip}     \fi
\ifx \showISBNxiii \undefined \def \showISBNxiii  #1{\unskip}     \fi
\ifx \showISSN     \undefined \def \showISSN      #1{\unskip}     \fi
\ifx \showLCCN     \undefined \def \showLCCN      #1{\unskip}     \fi
\ifx \shownote     \undefined \def \shownote      #1{#1}          \fi
\ifx \showarticletitle \undefined \def \showarticletitle #1{#1}   \fi
\ifx \showURL      \undefined \def \showURL       {\relax}        \fi
\providecommand\bibfield[2]{#2}
\providecommand\bibinfo[2]{#2}
\providecommand\natexlab[1]{#1}
\providecommand\showeprint[2][]{arXiv:#2}

\bibitem[Bai et~al\mbox{.}(2024)]%
        {bai2024}
\bibfield{author}{\bibinfo{person}{Songyuan Bai}, \bibinfo{person}{Hao Zheng}, \bibinfo{person}{Chen Tian}, \bibinfo{person}{Xiaoliang Wang}, \bibinfo{person}{Chang Liu}, \bibinfo{person}{Xin Jin}, \bibinfo{person}{Fu Xiao}, \bibinfo{person}{Qiao Xiang}, \bibinfo{person}{Wanchun Dou}, {and} \bibinfo{person}{Guihai Chen}.} \bibinfo{year}{2024}\natexlab{}.
\newblock \showarticletitle{{Unison}: {A} {Parallel-Efficient} and {User-Transparent} {Network} {Simulation} {Kernel}}. In \bibinfo{booktitle}{\emph{Proceedings of the Nineteenth European Conference on Computer Systems}} \emph{(\bibinfo{series}{EuroSys '24})}. \bibinfo{publisher}{Association for Computing Machinery}, \bibinfo{pages}{115–131}.
\newblock
\href{https://doi.org/10.1145/3627703.3629574}{doi:\nolinkurl{10.1145/3627703.3629574}}


\bibitem[Camara et~al\mbox{.}(2014)]%
        {camara2014}
\bibfield{author}{\bibinfo{person}{Daniel Camara}, \bibinfo{person}{Hajime Tazaki}, \bibinfo{person}{Emilio Mancini}, \bibinfo{person}{Thierry Turletti}, \bibinfo{person}{Walid Dabbous}, {and} \bibinfo{person}{Mathieu Lacage}.} \bibinfo{year}{2014}\natexlab{}.
\newblock \showarticletitle{{DCE}: {Test} the real code of your protocols and applications over simulated networks}.
\newblock \bibinfo{journal}{\emph{IEEE Communications Magazine}} \bibinfo{volume}{52}, \bibinfo{number}{3} (\bibinfo{year}{2014}), \bibinfo{pages}{104--110}.
\newblock
\href{https://doi.org/10.1109/MCOM.2014.6766093}{doi:\nolinkurl{10.1109/MCOM.2014.6766093}}


\bibitem[Contributors(2022)]%
        {mininet2022}
\bibfield{author}{\bibinfo{person}{Mininet~Project Contributors}.} \bibinfo{year}{2022}\natexlab{}.
\newblock \bibinfo{title}{{Mininet} -- {An} {Instant} {Virtual} {Network} on your {Laptop} (or other {PC})}.
\newblock
\urldef\tempurl%
\url{https://mininet.org/}
\showURL{%
\tempurl}


\bibitem[{ESnet / Lawrence Berkeley National Laboratory}(2019)]%
        {iperf3}
\bibfield{author}{\bibinfo{person}{{ESnet / Lawrence Berkeley National Laboratory}}.} \bibinfo{year}{2019}\natexlab{}.
\newblock \bibinfo{title}{{iPerf3}: {A} {TCP}, {UDP}, and {SCTP} network bandwidth measurement tool}.
\newblock \bibinfo{howpublished}{\url{https://github.com/esnet/iperf}}.
\newblock


\bibitem[{Fortune Business Insights}(2025)]%
        {fortune2025}
\bibfield{author}{\bibinfo{person}{{Fortune Business Insights}}.} \bibinfo{year}{2025}\natexlab{}.
\newblock \bibinfo{booktitle}{\emph{Satellite {Mega} {Constellations} {Market} {Size}, {Share}, {Growth} and {Industry} {Analysis}, {By} {Component}, {By} {Orbit} {Type}, {By} {Application}, and {Regional} {Forecast}, 2025-2032}}.
\newblock
\urldef\tempurl%
\url{https://www.fortunebusinessinsights.com/satellite-mega-constellations-market-112989}
\showURL{%
\tempurl}


\bibitem[Garcia et~al\mbox{.}(2023)]%
        {garcia2023}
\bibfield{author}{\bibinfo{person}{Johan Garcia}, \bibinfo{person}{Simon Sundberg}, \bibinfo{person}{Giuseppe Caso}, {and} \bibinfo{person}{Anna Brunstrom}.} \bibinfo{year}{2023}\natexlab{}.
\newblock \showarticletitle{Multi-Timescale Evaluation of Starlink Throughput}. In \bibinfo{booktitle}{\emph{Proceedings of the 1st ACM Workshop on LEO Networking and Communication}} (Madrid, Spain) \emph{(\bibinfo{series}{LEO-NET '23})}. \bibinfo{publisher}{Association for Computing Machinery}, \bibinfo{address}{New York, NY, USA}, \bibinfo{pages}{31–36}.
\newblock
\showISBNx{9798400703324}
\href{https://doi.org/10.1145/3614204.3616108}{doi:\nolinkurl{10.1145/3614204.3616108}}


\bibitem[Hemminger et~al\mbox{.}(2005)]%
        {hemminger2005}
\bibfield{author}{\bibinfo{person}{Stephen Hemminger} {et~al\mbox{.}}} \bibinfo{year}{2005}\natexlab{}.
\newblock \showarticletitle{{Network} emulation with {NetEm}}. In \bibinfo{booktitle}{\emph{Linux Conf AU}}, Vol.~\bibinfo{volume}{5}. \bibinfo{address}{Canberra, Australia}, \bibinfo{pages}{2005}.
\newblock


\bibitem[Hofstätter et~al\mbox{.}(2025)]%
        {hofstatter2025}
\bibfield{author}{\bibinfo{person}{Matthias Hofstätter}, \bibinfo{person}{Jörg Deutschmann}, \bibinfo{person}{Raffaello Secchi}, \bibinfo{person}{Gorry Fairhurst}, {and} \bibinfo{person}{Reinhard German}.} \bibinfo{year}{2025}\natexlab{}.
\newblock \showarticletitle{Careful {Resume}: {Design} and {Analysis} with {Picoquic} over {Satellite} {Paths}}. In \bibinfo{booktitle}{\emph{2025 12th Advanced Satellite Multimedia Systems Conference and the 18th Signal Processing for Space Communications Workshop (ASMS/SPSC)}}. \bibinfo{pages}{1--8}.
\newblock
\href{https://doi.org/10.1109/ASMS/SPSC64465.2025.10946055}{doi:\nolinkurl{10.1109/ASMS/SPSC64465.2025.10946055}}


\bibitem[INRIA(2013)]%
        {dce2013}
\bibfield{author}{\bibinfo{person}{INRIA}.} \bibinfo{year}{2013}\natexlab{}.
\newblock \bibinfo{title}{Direct {Code} {Execution} {(DCE)} {Manual}}.
\newblock \bibinfo{howpublished}{\url{https://www.nsnam.org/docs/dce/release/1.1/manual/singlehtml/index.html}}.
\newblock


\bibitem[Jiang et~al\mbox{.}(2023)]%
        {jiang2023}
\bibfield{author}{\bibinfo{person}{Weiwei Jiang}, \bibinfo{person}{Yafeng Zhan}, \bibinfo{person}{Xiaolong Xiao}, {and} \bibinfo{person}{Guanglin Sha}.} \bibinfo{year}{2023}\natexlab{}.
\newblock \showarticletitle{{Network} {Simulators} for {Satellite-Terrestrial} {Integrated} {Networks}: {A} {Survey}}.
\newblock \bibinfo{journal}{\emph{IEEE Access}}  \bibinfo{volume}{11} (\bibinfo{year}{2023}), \bibinfo{pages}{98269--98292}.
\newblock
\href{https://doi.org/10.1109/ACCESS.2023.3313229}{doi:\nolinkurl{10.1109/ACCESS.2023.3313229}}


\bibitem[Kassem and Sastry(2024)]%
        {kassem2024}
\bibfield{author}{\bibinfo{person}{Mohamed~M. Kassem} {and} \bibinfo{person}{Nishanth Sastry}.} \bibinfo{year}{2024}\natexlab{}.
\newblock \showarticletitle{$x$eoverse: {A} {Real-time} {Simulation} {Platform} for {Large} {LEO} {Satellite} {Mega-Constellations}}. In \bibinfo{booktitle}{\emph{2024 IFIP Networking Conference (IFIP Networking)}}. \bibinfo{pages}{1--9}.
\newblock
\href{https://doi.org/10.23919/IFIPNetworking62109.2024.10619898}{doi:\nolinkurl{10.23919/IFIPNetworking62109.2024.10619898}}


\bibitem[Kassing et~al\mbox{.}(2020)]%
        {kassing2020}
\bibfield{author}{\bibinfo{person}{Simon Kassing}, \bibinfo{person}{Debopam Bhattacherjee}, \bibinfo{person}{Andr\'{e}~Baptista \'{A}guas}, \bibinfo{person}{Jens~Eirik Saethre}, {and} \bibinfo{person}{Ankit Singla}.} \bibinfo{year}{2020}\natexlab{}.
\newblock \showarticletitle{{Exploring} the "{Internet} from space" with {Hypatia}}. In \bibinfo{booktitle}{\emph{Proceedings of the ACM Internet Measurement Conference}} \emph{(\bibinfo{series}{IMC '20})}. \bibinfo{publisher}{Association for Computing Machinery}, \bibinfo{address}{New York, NY, USA}, \bibinfo{pages}{214–229}.
\newblock
\showISBNx{9781450381383}
\href{https://doi.org/10.1145/3419394.3423635}{doi:\nolinkurl{10.1145/3419394.3423635}}


\bibitem[Kosek et~al\mbox{.}(2022)]%
        {kosek2022}
\bibfield{author}{\bibinfo{person}{Mike Kosek}, \bibinfo{person}{Hendrik Cech}, \bibinfo{person}{Vaibhav Bajpai}, {and} \bibinfo{person}{J{\"o}rg Ott}.} \bibinfo{year}{2022}\natexlab{}.
\newblock \showarticletitle{{Exploring} proxying {QUIC} and {HTTP/3} for satellite communication}. In \bibinfo{booktitle}{\emph{2022 IFIP Networking Conference (IFIP Networking)}}. IEEE, \bibinfo{pages}{1--9}.
\newblock
\urldef\tempurl%
\url{https://doi.org/10.23919/IFIPNetworking55013.2022.9829773}
\showURL{%
\tempurl}


\bibitem[Lai et~al\mbox{.}(2023)]%
        {lai2023}
\bibfield{author}{\bibinfo{person}{Zeqi Lai}, \bibinfo{person}{Hewu Li}, \bibinfo{person}{Yangtao Deng}, \bibinfo{person}{Qian Wu}, \bibinfo{person}{Jun Liu}, \bibinfo{person}{Yuanjie Li}, \bibinfo{person}{Jihao Li}, \bibinfo{person}{Lixin Liu}, \bibinfo{person}{Weisen Liu}, {and} \bibinfo{person}{Jianping Wu}.} \bibinfo{year}{2023}\natexlab{}.
\newblock \showarticletitle{StarryNet: empowering researchers to evaluate futuristic integrated space and terrestrial networks}. In \bibinfo{booktitle}{\emph{20th USENIX Symposium on Networked Systems Design and Implementation (NSDI 23)}}. \bibinfo{pages}{1309--1324}.
\newblock
\urldef\tempurl%
\url{https://www.usenix.org/conference/nsdi23/presentation/lai-zeqi}
\showURL{%
\tempurl}


\bibitem[Lai et~al\mbox{.}(2024)]%
        {lai2024}
\bibfield{author}{\bibinfo{person}{Zeqi Lai}, \bibinfo{person}{Zonglun Li}, \bibinfo{person}{Qian Wu}, \bibinfo{person}{Hewu Li}, \bibinfo{person}{Weisen Liu}, \bibinfo{person}{Yijie Liu}, \bibinfo{person}{Xin Xie}, \bibinfo{person}{Yuanjie Li}, {and} \bibinfo{person}{Jun Liu}.} \bibinfo{year}{2024}\natexlab{}.
\newblock \showarticletitle{Mind the {Misleading} {Effects} of {LEO} {Mobility} on {End-to-End} {Congestion} {Control}}. In \bibinfo{booktitle}{\emph{Proceedings of the 23rd ACM Workshop on Hot Topics in Networks}} (Irvine, CA, USA) \emph{(\bibinfo{series}{HotNets '24})}. \bibinfo{publisher}{Association for Computing Machinery}, \bibinfo{address}{New York, NY, USA}, \bibinfo{pages}{34–42}.
\newblock
\showISBNx{9798400712722}
\href{https://doi.org/10.1145/3696348.3696867}{doi:\nolinkurl{10.1145/3696348.3696867}}


\bibitem[Laniewski et~al\mbox{.}(2024)]%
        {laniewski2024}
\bibfield{author}{\bibinfo{person}{Dominic Laniewski}, \bibinfo{person}{Eric Lanfer}, \bibinfo{person}{Bernd Meijerink}, \bibinfo{person}{Roland van Rijswijk-Deij}, {and} \bibinfo{person}{Nils Aschenbruck}.} \bibinfo{year}{2024}\natexlab{}.
\newblock \showarticletitle{{WetLinks}: a large-scale longitudinal starlink dataset with contiguous weather data}.
\newblock \bibinfo{journal}{\emph{arXiv preprint arXiv:2402.16448}} (\bibinfo{year}{2024}).
\newblock
\href{https://doi.org/10.48550/arXiv.2402.16448}{doi:\nolinkurl{10.48550/arXiv.2402.16448}}


\bibitem[Liu(2022)]%
        {liu2022}
\bibfield{author}{\bibinfo{person}{Hangbin Liu}.} \bibinfo{year}{2022}\natexlab{}.
\newblock \bibinfo{booktitle}{\emph{{An} {Introduction} to {Linux} {Bridging} {Commands} and {Features} -- {Red} {Hat} {Developer}}}.
\newblock
\urldef\tempurl%
\url{https://developers.redhat.com/articles/2022/04/06/introduction-linux-bridging-commands-and-features}
\showURL{%
\tempurl}


\bibitem[Lu et~al\mbox{.}(2025)]%
        {lu2025}
\bibfield{author}{\bibinfo{person}{Wenhao Lu}, \bibinfo{person}{Zhiyuan Wang}, \bibinfo{person}{Hefan Zhang}, \bibinfo{person}{Shan Zhang}, {and} \bibinfo{person}{Hongbin Luo}.} \bibinfo{year}{2025}\natexlab{}.
\newblock \showarticletitle{OpenSN: {An} {Open} {Source} {Library} for {Emulating} {LEO} {Satellite} {Networks}}.
\newblock \bibinfo{journal}{\emph{IEEE Transactions on Parallel and Distributed Systems}} \bibinfo{volume}{36}, \bibinfo{number}{8} (\bibinfo{year}{2025}), \bibinfo{pages}{1574--1590}.
\newblock
\href{https://doi.org/10.1109/TPDS.2025.3575920}{doi:\nolinkurl{10.1109/TPDS.2025.3575920}}


\bibitem[Mohan et~al\mbox{.}(2024)]%
        {mohan2024}
\bibfield{author}{\bibinfo{person}{Nitinder Mohan}, \bibinfo{person}{Andrew~E. Ferguson}, \bibinfo{person}{Hendrik Cech}, \bibinfo{person}{Rohan Bose}, \bibinfo{person}{Prakita~Rayyan Renatin}, \bibinfo{person}{Mahesh~K. Marina}, {and} \bibinfo{person}{Jörg Ott}.} \bibinfo{year}{2024}\natexlab{}.
\newblock \showarticletitle{{A} {Multifaceted} {Look} at {Starlink} {Performance}}. In \bibinfo{booktitle}{\emph{Proceedings of the ACM Web Conference 2024}}. \bibinfo{publisher}{ACM}, \bibinfo{address}{Singapore, Singapore}, \bibinfo{pages}{2723–2734}.
\newblock
\href{https://doi.org/10.1145/3589334.3645328}{doi:\nolinkurl{10.1145/3589334.3645328}}


\bibitem[Niehoefer et~al\mbox{.}(2013)]%
        {niehoefer2013}
\bibfield{author}{\bibinfo{person}{Brian Niehoefer}, \bibinfo{person}{Sebastian Subik}, {and} \bibinfo{person}{Christian Wietfeld}.} \bibinfo{year}{2013}\natexlab{}.
\newblock \showarticletitle{The {CNI} {Open} {Source} {Satellite} {Simulator} based on {OMNeT++}}. \bibinfo{publisher}{ACM}.
\newblock
\href{https://doi.org/10.4108/icst.simutools.2013.251580}{doi:\nolinkurl{10.4108/icst.simutools.2013.251580}}


\bibitem[nsnam(2017)]%
        {nsnam2017}
\bibfield{author}{\bibinfo{person}{nsnam}.} \bibinfo{year}{2017}\natexlab{}.
\newblock \bibinfo{title}{ns-3 {W}iki -- {HOWTO} make ns-3 interact with the real world}.
\newblock \bibinfo{howpublished}{\url{https://www.nsnam.org/wiki/HOWTO_make_ns-3_interact_with_the_real_world}}.
\newblock


\bibitem[nsnam(2025a)]%
        {nsnam2025-1}
\bibfield{author}{\bibinfo{person}{nsnam}.} \bibinfo{year}{2025}\natexlab{a}.
\newblock \bibinfo{title}{ns-3 {Manual} -- 2.6 {RealTime}}.
\newblock \bibinfo{howpublished}{\url{https://www.nsnam.org/docs/manual/html/realtime.html}}.
\newblock


\bibitem[nsnam(2025b)]%
        {nsnam2025-2}
\bibfield{author}{\bibinfo{person}{nsnam}.} \bibinfo{year}{2025}\natexlab{b}.
\newblock \bibinfo{title}{ns-3 {Manual} -- 3.1 {Random Variables}}.
\newblock \bibinfo{howpublished}{\url{https://www.nsnam.org/docs/manual/html/random-variables.html}}.
\newblock


\bibitem[Ohs et~al\mbox{.}(2025)]%
        {ohs2025}
\bibfield{author}{\bibinfo{person}{Robin Ohs}, \bibinfo{person}{Gregory~F. Stock}, \bibinfo{person}{Juan~A. Fraire}, \bibinfo{person}{Holger Hermanns}, {and} \bibinfo{person}{Andreas Schmidt}.} \bibinfo{year}{2025}\natexlab{}.
\newblock \showarticletitle{{PhantomLink}: {Emulating} {Virtual} {End-to-End} {Links} on {Ground} and in {Orbit}}. In \bibinfo{booktitle}{\emph{Proceedings of the 2025 Applied Networking Research Workshop}} (Madrid, Spain) \emph{(\bibinfo{series}{ANRW '25})}. \bibinfo{publisher}{Association for Computing Machinery}, \bibinfo{address}{New York, NY, USA}, \bibinfo{pages}{39–46}.
\newblock
\showISBNx{9798400720093}
\href{https://doi.org/10.1145/3744200.3744758}{doi:\nolinkurl{10.1145/3744200.3744758}}


\bibitem[Ottens et~al\mbox{.}(2025a)]%
        {ottens2025-1}
\bibfield{author}{\bibinfo{person}{Martin Ottens}, \bibinfo{person}{Jörg Deutschmann}, \bibinfo{person}{Kai-Steffen Hielscher}, {and} \bibinfo{person}{Reinhard German}.} \bibinfo{year}{2025}\natexlab{a}.
\newblock \showarticletitle{{Performance} {Evaluation} of ns-3 {Real-Time} {Emulation}}.
\newblock \bibinfo{journal}{\emph{IEEE Access}} (\bibinfo{year}{2025}).
\newblock
\href{https://doi.org/10.1109/ACCESS.2025.3555478}{doi:\nolinkurl{10.1109/ACCESS.2025.3555478}}


\bibitem[Ottens et~al\mbox{.}(2025b)]%
        {ottens2025-3}
\bibfield{author}{\bibinfo{person}{Martin Ottens}, \bibinfo{person}{Jörg Deutschmann}, \bibinfo{person}{Kai-Steffen Hielscher}, {and} \bibinfo{person}{Reinhard German}.} \bibinfo{year}{2025}\natexlab{b}.
\newblock \showarticletitle{{Proto2Testbed}: {Towards} an {Integrated} {Testbed} for {Evaluating} {End-to-End} {Security} {Protocols} in {Satellite} {Constellations}}. In \bibinfo{booktitle}{\emph{2025 12th Advanced Satellite Multimedia Systems Conference and the 18th Signal Processing for Space Communications Workshop (ASMS/SPSC)}}. \bibinfo{pages}{1--8}.
\newblock
\href{https://doi.org/10.1109/ASMS/SPSC64465.2025.10946051}{doi:\nolinkurl{10.1109/ASMS/SPSC64465.2025.10946051}}


\bibitem[Ottens et~al\mbox{.}(2025c)]%
        {ottens2025-2}
\bibfield{author}{\bibinfo{person}{Martin Ottens}, \bibinfo{person}{Kai-Steffen Hielscher}, {and} \bibinfo{person}{Reinhard German}.} \bibinfo{year}{2025}\natexlab{c}.
\newblock \showarticletitle{From {Simulation} to {Emulation}: {A} {System} {Design} for {Real-Time} {Replay} of {Simulated} {Network} {Path} {Characteristics}}. In \bibinfo{booktitle}{\emph{Proceedings of the 2025 International Conference on Ns-3}} (Osaka, JP) \emph{(\bibinfo{series}{ICNS3 '25})}. \bibinfo{publisher}{Association for Computing Machinery}, \bibinfo{address}{New York, NY, USA}, \bibinfo{pages}{62–69}.
\newblock
\showISBNx{9798400715174}
\href{https://doi.org/10.1145/3747204.3747206}{doi:\nolinkurl{10.1145/3747204.3747206}}


\bibitem[Pan et~al\mbox{.}(2022)]%
        {pan2022}
\bibfield{author}{\bibinfo{person}{Sibo Pan}, \bibinfo{person}{Ruifeng Gao}, \bibinfo{person}{Yingdong Hu}, {and} \bibinfo{person}{Ye Li}.} \bibinfo{year}{2022}\natexlab{}.
\newblock \showarticletitle{{TCP} {Performance} in {Satellite} {Backhauling} for {Maritime} {Communications}: {A} ns-3 {Study}}. In \bibinfo{booktitle}{\emph{2022 14th International Conference on Wireless Communications and Signal Processing (WCSP)}}. \bibinfo{pages}{877--882}.
\newblock
\href{https://doi.org/10.1109/WCSP55476.2022.10039046}{doi:\nolinkurl{10.1109/WCSP55476.2022.10039046}}


\bibitem[Prakash and Abdrabou(2020)]%
        {prakash2020}
\bibfield{author}{\bibinfo{person}{Monika Prakash} {and} \bibinfo{person}{Atef Abdrabou}.} \bibinfo{year}{2020}\natexlab{}.
\newblock \showarticletitle{On the {Fidelity} of {NS-3} {Simulations} of {Wireless} {Multipath} {TCP} {Connections}}.
\newblock \bibinfo{journal}{\emph{Sensors}} \bibinfo{volume}{20}, \bibinfo{number}{24} (\bibinfo{year}{2020}).
\newblock
\showISSN{1424-8220}
\href{https://doi.org/10.3390/s20247289}{doi:\nolinkurl{10.3390/s20247289}}


\bibitem[Ramakrishnan et~al\mbox{.}(2001)]%
        {rfc3168}
\bibfield{author}{\bibinfo{person}{K. Ramakrishnan}, \bibinfo{person}{S. Floyd}, {and} \bibinfo{person}{David~L. Black}.} \bibinfo{year}{2001}\natexlab{}.
\newblock \bibinfo{title}{The {Addition} of {Explicit} {Congestion} {Notification} {(ECN)} to {IP}}.
\newblock \bibinfo{howpublished}{RFC 3168}.
\newblock
\href{https://doi.org/10.17487/RFC3168}{doi:\nolinkurl{10.17487/RFC3168}}


\bibitem[Roth et~al\mbox{.}(2022)]%
        {roth2022}
\bibfield{author}{\bibinfo{person}{Manuel M.~H. Roth}, \bibinfo{person}{Hartmut Brandt}, {and} \bibinfo{person}{Hermann Bischl}.} \bibinfo{year}{2022}\natexlab{}.
\newblock \showarticletitle{{Distributed} {SDN-based} {Load-balanced} {Routing} for {Low} {Earth} {Orbit} {Satellite} {Constellation} {Networks}}. In \bibinfo{booktitle}{\emph{2022 11th Advanced Satellite Multimedia Systems Conference and the 17th Signal Processing for Space Communications Workshop (ASMS/SPSC)}}. \bibinfo{pages}{1--8}.
\newblock
\href{https://doi.org/10.1109/ASMS/SPSC55670.2022.9914690}{doi:\nolinkurl{10.1109/ASMS/SPSC55670.2022.9914690}}


\bibitem[Ruan et~al\mbox{.}(2025)]%
        {ruan2025}
\bibfield{author}{\bibinfo{person}{Guohao Ruan}, \bibinfo{person}{Tian Pan}, \bibinfo{person}{Haibin Song}, \bibinfo{person}{Qiang Fu}, \bibinfo{person}{Haonan Li}, \bibinfo{person}{Yi Liu}, \bibinfo{person}{Wenxuan Zhao}, \bibinfo{person}{Jun Yao}, {and} \bibinfo{person}{Tao Huang}.} \bibinfo{year}{2025}\natexlab{}.
\newblock \showarticletitle{{NSDocker}: {A} {Lightweight} and {Realistic} {Satellite} {Network} {Emulator} {Integrating} {NS-3} and {Docker}}. In \bibinfo{booktitle}{\emph{Proceedings of the 9th Asia-Pacific Workshop on Networking}} \emph{(\bibinfo{series}{APNET '25})}. \bibinfo{publisher}{Association for Computing Machinery}, \bibinfo{address}{New York, NY, USA}, \bibinfo{pages}{291–293}.
\newblock
\showISBNx{9798400714016}
\href{https://doi.org/10.1145/3735358.3737767}{doi:\nolinkurl{10.1145/3735358.3737767}}


\bibitem[Schubert et~al\mbox{.}(2022)]%
        {schubert2022}
\bibfield{author}{\bibinfo{person}{Tim Schubert}, \bibinfo{person}{Lars Wolf}, {and} \bibinfo{person}{Ulf Kulau}.} \bibinfo{year}{2022}\natexlab{}.
\newblock \showarticletitle{ns-3-leo: {Evaluation} {Tool} for {Satellite} {Swarm} {Communication} {Protocols}}.
\newblock \bibinfo{journal}{\emph{IEEE Access}}  \bibinfo{volume}{10} (\bibinfo{year}{2022}), \bibinfo{pages}{11527--11537}.
\newblock
\href{https://doi.org/10.1109/ACCESS.2022.3146770}{doi:\nolinkurl{10.1109/ACCESS.2022.3146770}}


\bibitem[Song et~al\mbox{.}(2024)]%
        {song2024}
\bibfield{author}{\bibinfo{person}{Haibin Song}, \bibinfo{person}{Tian Pan}, \bibinfo{person}{Guohao Ruan}, \bibinfo{person}{Yan Zheng}, \bibinfo{person}{Ying Wan}, \bibinfo{person}{Jiao Zhang}, \bibinfo{person}{Tao Huang}, {and} \bibinfo{person}{Yunjie Liu}.} \bibinfo{year}{2024}\natexlab{}.
\newblock \showarticletitle{Accelerating {Mega-Scale} {Satellite} {Network} {Simulation} in {NS-3} via {MPI-based} {Parallelization}}. In \bibinfo{booktitle}{\emph{ICC 2024 - IEEE International Conference on Communications}}. \bibinfo{pages}{5610--5615}.
\newblock
\href{https://doi.org/10.1109/ICC51166.2024.10622934}{doi:\nolinkurl{10.1109/ICC51166.2024.10622934}}


\bibitem[Tian et~al\mbox{.}(2024)]%
        {tian2024}
\bibfield{author}{\bibinfo{person}{Weibiao Tian}, \bibinfo{person}{Ye Li}, \bibinfo{person}{Jinwei Zhao}, \bibinfo{person}{Sheng Wu}, {and} \bibinfo{person}{Jianping Pan}.} \bibinfo{year}{2024}\natexlab{}.
\newblock \showarticletitle{An {eBPF-Based} {Trace-Driven} {Emulation} {Method} for {Satellite} {Networks}}.
\newblock \bibinfo{journal}{\emph{IEEE Networking Letters}} \bibinfo{volume}{6}, \bibinfo{number}{3} (\bibinfo{year}{2024}), \bibinfo{pages}{188--192}.
\newblock
\href{https://doi.org/10.1109/LNET.2024.3472034}{doi:\nolinkurl{10.1109/LNET.2024.3472034}}


\bibitem[Wehrle et~al\mbox{.}(2010)]%
        {wehrle2010}
\bibfield{author}{\bibinfo{person}{Klaus Wehrle}, \bibinfo{person}{Mesut G{\"u}nes}, {and} \bibinfo{person}{James Gross}.} \bibinfo{year}{2010}\natexlab{}.
\newblock \bibinfo{booktitle}{\emph{Modeling and tools for network simulation}}.
\newblock \bibinfo{publisher}{Springer Science \& Business Media}.
\newblock


\end{thebibliography}

\clearpage

\begin{minipage}{\textwidth}%
  \appendix
  \raggedbottom
  \section{EXAMPLE TRACE FILE EXCERPT}
\begin{lstlisting}[language=csv,firstnumber=1,caption=Excerpt from the Trace File of the forward path with updates of the path characteristics in 10~ms intervals in \ac{csv} format. The file was generated with the Kuiper constellation from Scenario~1 (Section~\ref{fig:scenario_1_trace}) between \acp{gs} in Boston and Paris. The \ac{bdp} is not included in the \texttt{queue\_capacity} field.,label=lst:trace_file]
  keep_us, <@\textcolor{DarkOrchid}{delay\_us,}@> <@\textcolor{WildStrawberry}{stddev\_us,}@> <@\textcolor{Green}{min\_link\_bps,}@> <@\textcolor{Red}{max\_link\_bps,}@> <@\textcolor{RedOrange}{queue\_capacity,}@> <@\textcolor{RoyalBlue}{hops,}@> dropratio, <@\textcolor{DarkOrchid}{route\_id}@>
  ...
  10000,   <@\textcolor{DarkOrchid}{22412,}@>    <@\textcolor{WildStrawberry}{341.92,}@>   <@\textcolor{Green}{36662240.00,}@>  <@\textcolor{Red}{49519360.00,}@>  <@\textcolor{RedOrange}{199,}@>          <@\textcolor{RoyalBlue}{6,}@>    0.00,      <@\textcolor{DarkOrchid}{16}@>
  10000,   <@\textcolor{DarkOrchid}{21919,}@>    <@\textcolor{WildStrawberry}{179.09,}@>   <@\textcolor{Green}{35821120.00,}@>  <@\textcolor{Red}{49519360.00,}@>  <@\textcolor{RedOrange}{200,}@>          <@\textcolor{RoyalBlue}{6,}@>    0.00,      <@\textcolor{DarkOrchid}{16}@>
  10000,   <@\textcolor{DarkOrchid}{22165,}@>    <@\textcolor{WildStrawberry}{281.26,}@>   <@\textcolor{Green}{35580800.00,}@>  <@\textcolor{Red}{49519360.00,}@>  <@\textcolor{RedOrange}{000,}@>          <@\textcolor{RoyalBlue}{6,}@>    0.20,      <@\textcolor{DarkOrchid}{16}@>
  10000,   <@\textcolor{DarkOrchid}{00000,}@>    <@\textcolor{WildStrawberry}{000.00,}@>   <@\textcolor{Green}{00000000.00,}@>  <@\textcolor{Red}{00000000.00,}@>  <@\textcolor{RedOrange}{000,}@>          <@\textcolor{RoyalBlue}{0,}@>    1.00,      <@\textcolor{DarkOrchid}{00}@>
  10000,   <@\textcolor{DarkOrchid}{00000,}@>    <@\textcolor{WildStrawberry}{000.00,}@>   <@\textcolor{Green}{00000000.00,}@>  <@\textcolor{Red}{00000000.00,}@>  <@\textcolor{RedOrange}{000,}@>          <@\textcolor{RoyalBlue}{0,}@>    1.00,      <@\textcolor{DarkOrchid}{00}@>
  ...      
  10000,   <@\textcolor{DarkOrchid}{00000,}@>    <@\textcolor{WildStrawberry}{000.00,}@>   <@\textcolor{Green}{00000000.00,}@>  <@\textcolor{Red}{00000000.00,}@>  <@\textcolor{RedOrange}{000,}@>          <@\textcolor{RoyalBlue}{0,}@>    1.00,      <@\textcolor{DarkOrchid}{00}@>
  10000,   <@\textcolor{DarkOrchid}{00000,}@>    <@\textcolor{WildStrawberry}{000.00,}@>   <@\textcolor{Green}{00000000.00,}@>  <@\textcolor{Red}{00000000.00,}@>  <@\textcolor{RedOrange}{000,}@>          <@\textcolor{RoyalBlue}{0,}@>    1.00,      <@\textcolor{DarkOrchid}{00}@>
  10000,   <@\textcolor{DarkOrchid}{23930,}@>    <@\textcolor{WildStrawberry}{000.00,}@>   <@\textcolor{Green}{40026720.00,}@>  <@\textcolor{Red}{47476640.00,}@>  <@\textcolor{RedOrange}{040,}@>          <@\textcolor{RoyalBlue}{6,}@>    0.80,      <@\textcolor{DarkOrchid}{18}@>
  10000,   <@\textcolor{DarkOrchid}{23799,}@>    <@\textcolor{WildStrawberry}{416.39,}@>   <@\textcolor{Green}{40146880.00,}@>  <@\textcolor{Red}{47356480.00,}@>  <@\textcolor{RedOrange}{200,}@>          <@\textcolor{RoyalBlue}{6,}@>    0.00,      <@\textcolor{DarkOrchid}{18}@>
  10000,   <@\textcolor{DarkOrchid}{23737,}@>    <@\textcolor{WildStrawberry}{311.80,}@>   <@\textcolor{Green}{39425920.00,}@>  <@\textcolor{Red}{47236320.00,}@>  <@\textcolor{RedOrange}{200,}@>          <@\textcolor{RoyalBlue}{6,}@>    0.00,      <@\textcolor{DarkOrchid}{18}@>
  ...
\end{lstlisting}
\end{minipage}

\end{document}